\documentclass[12pt]{iopart}
\usepackage{graphics,iopams,mathrsfs}

\makeatletter
\def\@fnsymbol#1{^{\thefootnote}\relax}
\makeatother

\begin{document}

\title[Optimized Effective Potential Model for Double Perovskites]
{Optimized Effective Potential Model for the Double Perovskites
Sr$_{2-x}$Y$_x$VMoO$_6$ and Sr$_{2-x}$Y$_x$VTcO$_6$}

\author{I V Solovyev}

\address{
National Institute for Materials Science,
1-2-1 Sengen, Tsukuba, Ibaraki 305-0047, Japan}
\ead{solovyev.igor@nims.go.jp}

\begin{abstract}
In attempt to explore half-metallic properties of the
double perovskites Sr$_{2-x}$Y$_x$VMoO$_6$ and Sr$_{2-x}$Y$_x$VTcO$_6$,
we construct an effective low-energy model,
which describes the behavior of the $t_{2g}$-states of these compounds.
All parameters of such model are derived rigorously on the basis of
first-principles electronic structure calculations. In order to solve this model
we employ the optimized effective potential method and treat the correlation
interactions in the random phase approximation. Although correlation interactions
considerably
reduce the intraatomic exchange splitting in comparison with the
Hartree-Fock method,
this splitting still substantially exceeds the typical values
obtained in the local-spin-density approximation (LSDA),
which alters many predictions based on the LSDA. Our main results are summarized
as follows: (i) all ferromagnetic states are expected to be half-metallic.
However, their energies are generally higher than
those of
the ferrimagnetic ordering between V- and Mo/Tc-sites
(except Sr$_2$VMoO$_6$); (ii) all ferrimagnetic states
are metallic (except fully insulating Y$_2$VTcO$_6$) and no half-metallic
antiferromagnetism has been found; (iii) moreover,
many of the
ferrimagnetic structures
appear to be unstable with respect to the spin-spiral alignment.
Thus, the true magnetic ground state of the most of these systems is
expected to be more complex.
In addition, we discuss several methodological issues related to the nonuniqueness of the
effective potential for the magnetic half-metallic and insulating states.
\end{abstract}


\maketitle

\section{Introduction}

  Recently double perovskites $A_{2-x}A'_xBB'$O$_6$, where
$A$ and $A'$ are the di- and three-valent (typically alkali-earth or rare-earth) elements, respectively,
and $B$ ($B'$) are the transition-metal elements,
have attracted considerable attention~\cite{DPreview},
following experimental discoveries of the room-temperature ferromagnetism
and the colossal-magnetoresistance effect in Sr$_2$FeMoO$_6$~\cite{Kobayashi},
which make these materials promising in the field of spin electronics.
The extraordinary properties of the double perovskites are
frequently attributed to their half-metallic electronic structure,
although the situation can be rather subtle~\cite{PRB02}.
Nevertheless, the first discoveries spark the new research activity
aiming at the systematic search of half-metallic double-perovskite materials.
Many such predictions
has been made on the basis of first-principles electronic structure calculations.
Besides ferromagnetic (FM) compounds, a particular attention was paid to the
half-metallic antiferromagnets -- the systems, where the
half-metallicity coexists with the zero net magnetization~\cite{vanLeuken,Pickett}.
However, even apart from the technological issues, related to the defects in the
double-perovskite structure,
there is a number of fundamental problems,
which call many predictions into question:
\begin{enumerate}
\item
Most of the calculations are based on the local-spin-density approximation (LSDA)
or its refinement -- the generalized gradient approximation, which have many limitations for the transition-metal oxides.
Particularly, all these theories employ the functional dependence of the
exchange-correlation energy obtained in the limit of homogeneous electron gas.
However, many unique properties of the double perovskites are related to the
alternation of two different transition-metal sites $B$ and $B'$, and in this sense
the systems are essentially inhomogeneous. Thus, even though the double-perovskite compounds
may be classified as itinerant electron magnets, it does not necessary mean that
their properties are well described within LSDA;
\item
In the vast majority of calculations, it is believed that the LSDA$+$$U$
technique, which incorporates the physics of on-site
Coulomb interactions~\cite{AZA,PRB94b},
can serve as a reasonable alternative to LSDA and provide more superior
description for the properties of double perovskites. However, the LSDA$+$$U$ scheme for the double perovskites
is ill-defined: there are many uncertainties related to the
values of the
on-site Coulomb repulsion $U$ at the sites $B$ and $B'$, which
are frequently taken as adjustable parameters, the subspace of ``localized orbitals'', and the double-counting term~\cite{PRB98}.
Moreover, the total energy in LSDA$+$$U$ is taken in the \textit{ad hoc}
form, which is valid only in the atomic limit and equivalent to some constrained
density-functional theory~\cite{PRB94b}.
In such a situation, all predictions of the half-metallic behavior
for the double perovskites
should be taken very cautiously.
\end{enumerate}

  In this paper we will illustrate how these problem could be resolved by
applying the optimized effective potential (OEP) method, and what the
consequences on the electronic and magnetic properties of the double perovskites
could be. The OEP method is formulated in the spirit of Kohn-Sham
density functional theory~\cite{KohnSham}, by introducing an auxiliary
one-electron problem, which is specified by some local effective potential
and which is used for the search of the true spin density through the minimization of
the total energy of the system~\cite{TalmanShadwick,KotaniAkai,Kotani,EngelSchmid}.
In this work we will apply the OEP technique for the solution of the low-energy model,
where all the parameters are derived rigorously, on the basis of 
first-principles electronic structure calculations.
Besides simplicity, such a model analysis typically allows one to concentrate
directly
on the microscopic picture, underlying the considered phenomena.
For the first applications of the OEP model,
it is more convenient to consider the
$t_{2g}$-systems, whose electronic and magnetic properties
are predetermined by the behavior of the $t_{2g}$-states of the sites
$B$ and $B'$ located near the Fermi level. In this case, the effective potential is specified by only three parameters,
which should be used for the minimization of the total energy.

\begin{figure}[h!]
\begin{center}
\resizebox{6cm}{!}{\includegraphics{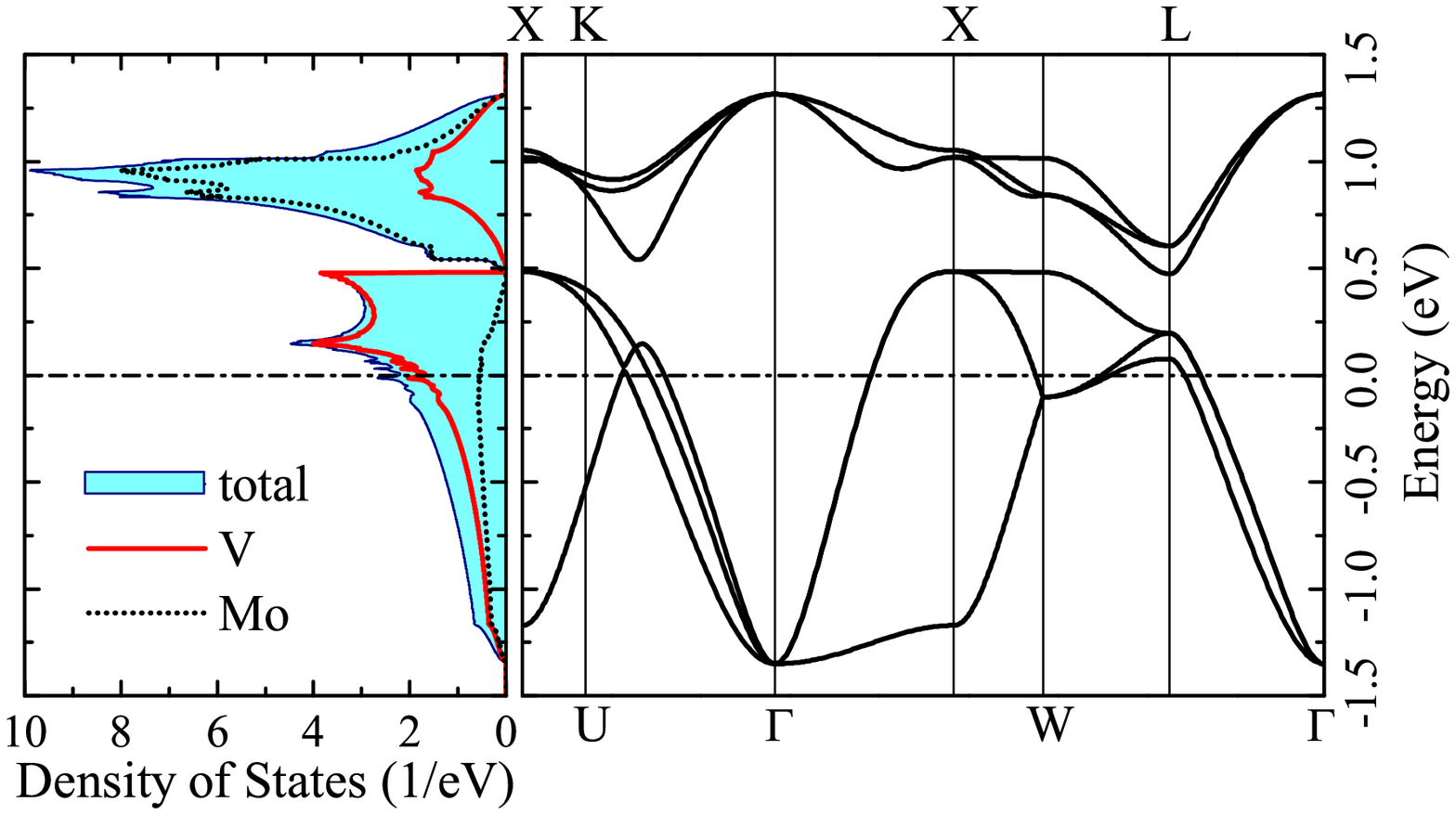}}
\resizebox{6cm}{!}{\includegraphics{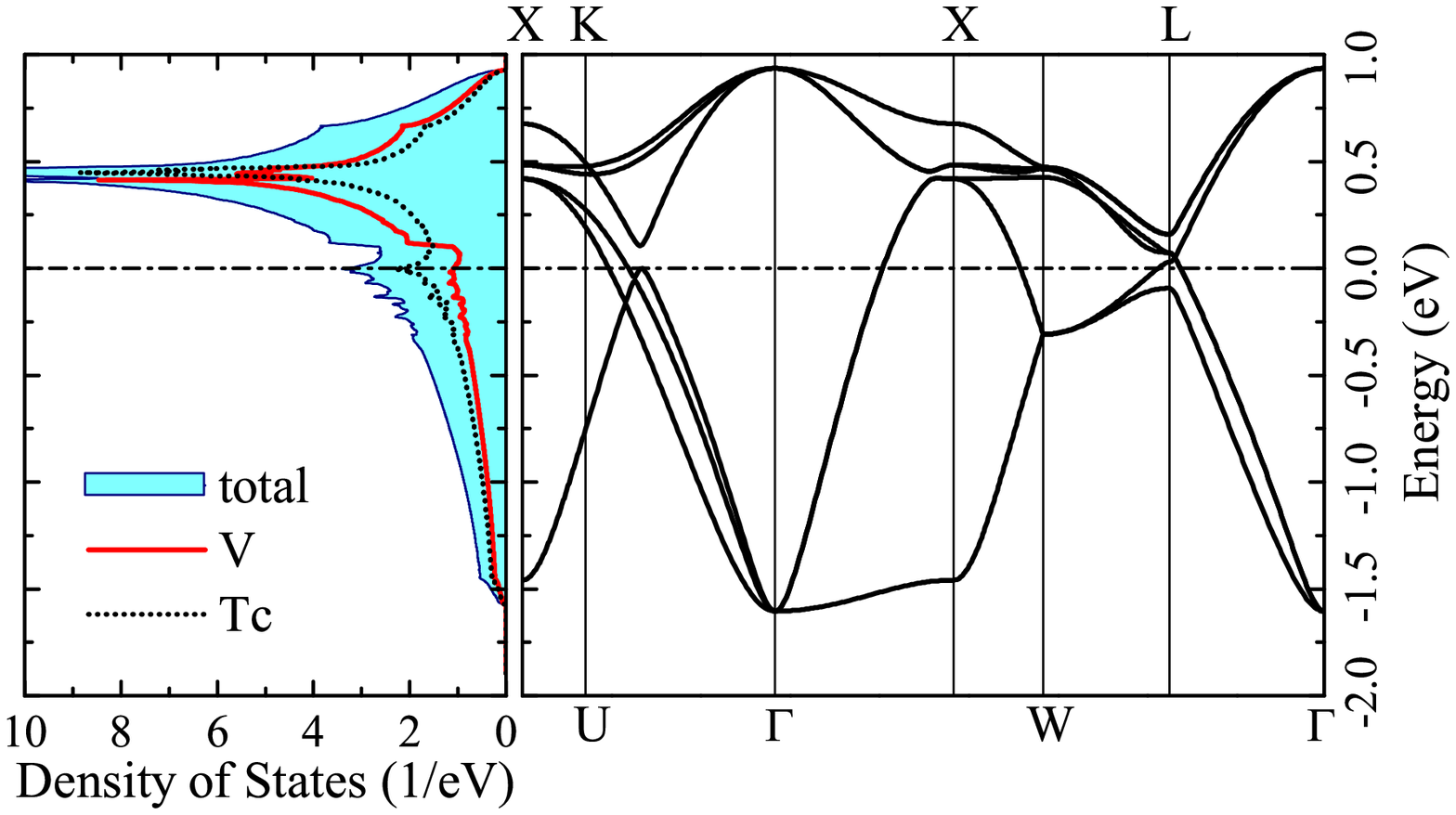}}
\end{center}
\caption{\label{fig.LDA} Electronic structure 
near the Fermi level
in the local-density approximation
(total and partial densities of states, and the energy dispersion of the $t_{2g}$-bands)
for Sr$_2$VMoO$_6$ (left) and Sr$_2$VTcO$_6$ (right). The position of the Fermi
level is shown by the dot-dashed line.}
\end{figure}

  We select two prototypical examples
Sr$_{2-x}$Y$_x$VMoO$_6$ and Sr$_{2-x}$Y$_x$VTcO$_6$,
for which we adopt the undistorted double perovskite structure with the
lattice parameters
$a =$ 7.846 and 7.899 \AA, respectively. In fact, $a =$ 7.846 \AA~ is the
experimental lattice constant for SrLaVMoO$_6$~\cite{Gotoh},
which is close to theoretical values obtained in the first-principles calculations~\cite{ParkMin}, and
$a =$ 7.899 \AA~ is the theoretical lattice constant for La$_2$VTcO$_6$~\cite{Wang}.
Deviations from the ideal double perovskite structure
are expected to be small~\cite{Wang} and
were not considered here.
The band-filling dependence due to the substitution of Sr by Y was
treated in the virtual-crystal approximation.
According to the electronic structure
calculations in the local-density approximation (LDA),
the basic difference between Sr$_2$VMoO$_6$ and Sr$_2$VTcO$_6$
is in the relative position of the V- and Mo(Tc)-states: in Sr$_2$VMoO$_6$,
the V- and Mo-states are separated in energy and form two nonoverlapping
bands, while in Sr$_2$VTcO$_6$ these states form one common band (Fig.~\ref{fig.LDA}).
Both Sr$_{2-x}$Y$_x$VMoO$_6$ and Sr$_{2-x}$Y$_x$VTcO$_6$
(and other isoelectronic to them compounds compounds)
were intensively discussed
in the literature in the context of the half-metallic applications. For instance,
SrLaVRuO$_6$, SrLaVMoO$_6$, and La$_2$VTcO$_6$ were proposed to be the
half-metallic antiferromagnets~\cite{Gotoh,Wang,Park,Uehara}.
On the other hand, the metallic ferrimagnetic (i.e., non spin-compensated) behavior
was also suggested for SrLaVMoO$_6$~\cite{ParkMin,Song}. In general,
such conclusions are sensitive to the values of the
on-site Coulomb repulsion $U$, which are used in the calculations.

  The rest of the paper is organized as follows. The method is discussed in Sec.~\ref{sec:method}.
Particularly, in Sec.~\ref{sec:LEmodel} we briefly describe construction of the
low-energy model for the double perovskites and in Sec.~\ref{sec:OEP} -- the basic
idea of the OEP method. Results of the model OEP calculations are shown in
Sec.~\ref{sec:results}. Particularly, the obtained electronic band structure
is discussed in Sec.~\ref{sec:DOS}, the screening of on-site
interactions due to the correlation effects and its consequences on the properties
of Sr$_{2-x}$Y$_x$VMoO$_6$ and Sr$_{2-x}$Y$_x$VTcO$_6$ is considered in
Sec.~\ref{sec:StonerI}, and the behavior of the total energy and stability of
the ferrimagnetic states are discussed in Sec.~\ref{sec:Etot}.
Finally, brief summary of the work is presented in Sec.~\ref{sec:summary}.

\section{\label{sec:method} Method}
\subsection{\label{sec:LEmodel} Construction and parameters of the low-energy model}
  The first step of our approach is the construction of the effective low-energy model
for the $t_{2g}$-states of the double perovskites:
\begin{equation}
\fl
\hat{\cal{H}}  =  \sum_{ij} \sum_\sigma \sum_{m m'}
t_{ij}^{m m'}\hat{c}^\dagger_{i m \sigma}
\hat{c}^{\phantom{\dagger}}_{j m' \sigma} +
  \frac{1}{2}
\sum_{i}  \sum_{\sigma \sigma'} \sum_{m m' m'' m'''} U^i_{m m' m'' m'''}
\hat{c}^\dagger_{i m \sigma} \hat{c}^\dagger_{i m'' \sigma'}
\hat{c}^{\phantom{\dagger}}_{i m' \sigma}
\hat{c}^{\phantom{\dagger}}_{i m''' \sigma'},
\label{eqn.ManyBodyH}
\end{equation}
which is derived in the Wannier basis, by starting
from the LDA band structure.
The Wannier functions are specified by the
indices $m$$=$ $xy$, $yz$, or $zx$, and $\sigma$ stands for the
spin indices $\uparrow$ or $\downarrow$.
The one-electron part $t_{ij}^{m m'}$ of the model Hamiltonian was
derived by using downfolding method, and the Coulomb (and exchange) interactions
$U^i_{m m' m'' m'''}$ were obtained by combining
the constrained density-functional theory with
the random-phase approximation (RPA).
The method was discussed in the literature, and for details the reader is
referred to the review article \cite{review2008}. There are three
sets of parameters in the model Hamiltonian (\ref{eqn.ManyBodyH}):
\begin{enumerate}
\item
the ``charge transfer'' energies, or the energy splitting between
``atomic'' $t_{2g}$-levels of the V and Mo (Tc) sites, which are described by
the site-diagonal part of $t_{ij}^{m m'}$.
The values of these parameters are $-$$391$ and 94 meV for the VMo and VTc compounds, respectively.
Thus, as expected from the LDA band structure (Fig.~\ref{fig.LDA}),
for the VMo compounds, the atomic V-states are located lower in energy and
well separated from the Mo-states, while for the
VTc compounds, the situation is reversal
and the splitting is considerably smaller;
\item
the transfer integrals, which are described by the off-diagonal elements of $t_{ij}^{m m'}$
with respect to the site indices. As expected, the largest parameter is the $dd\pi$-transfer 
integral between nearest-neighbor V- and Mo(Tc)-sites, which is $271$ (245) meV.
Another important parameter is the $dd\sigma$-transfer integral between
next nearest neighbors of the same atomic type. Its values 
in the V- and Mo-sublattices of the VMo compounds
are
$-$$72$ and $-$$116$ meV, respectively. Similar values were obtained for the VTc
compounds:
$-$$73$ and $-$$113$ meV for the V- and Tc-sublattice, respectively;
\item
intraatomic Coulomb and exchange interactions in the $t_{2g}$-band, which
are screened by all other states.
At each site of the system, these interactions
are
fully specified by two Kanamori parameters~\cite{Kanamori}: the intraorbital
Coulomb interaction $\mathcal{U} = U_{mmmm}$ and the exchange
interaction $\mathcal{J} = U_{mm'm'm}$ for $m$$\ne$$m'$.
In the perfect cubic environment, the third Kanamori parameter
$\mathcal{U}' = U_{mmm'm'}$
(the interorbital Coulomb interaction)
can be related to $~\mathcal{U}$ and $\mathcal{J}$ as
$~\mathcal{U}' = \mathcal{U} - 2 \mathcal{J}$~\cite{review2008}.
Other types of matrix elements of $\| U_{m m' m'' m'''} \|$ are equal to zero.
The band-filling dependence of $\mathcal{U}$, $\mathcal{U}'$, and $\mathcal{J}$,
as obtained in the virtual crystal approximation, is explained in Fig.~\ref{fig.UJ}.
\begin{figure}[h!]
\begin{center}
\resizebox{6cm}{!}{\includegraphics{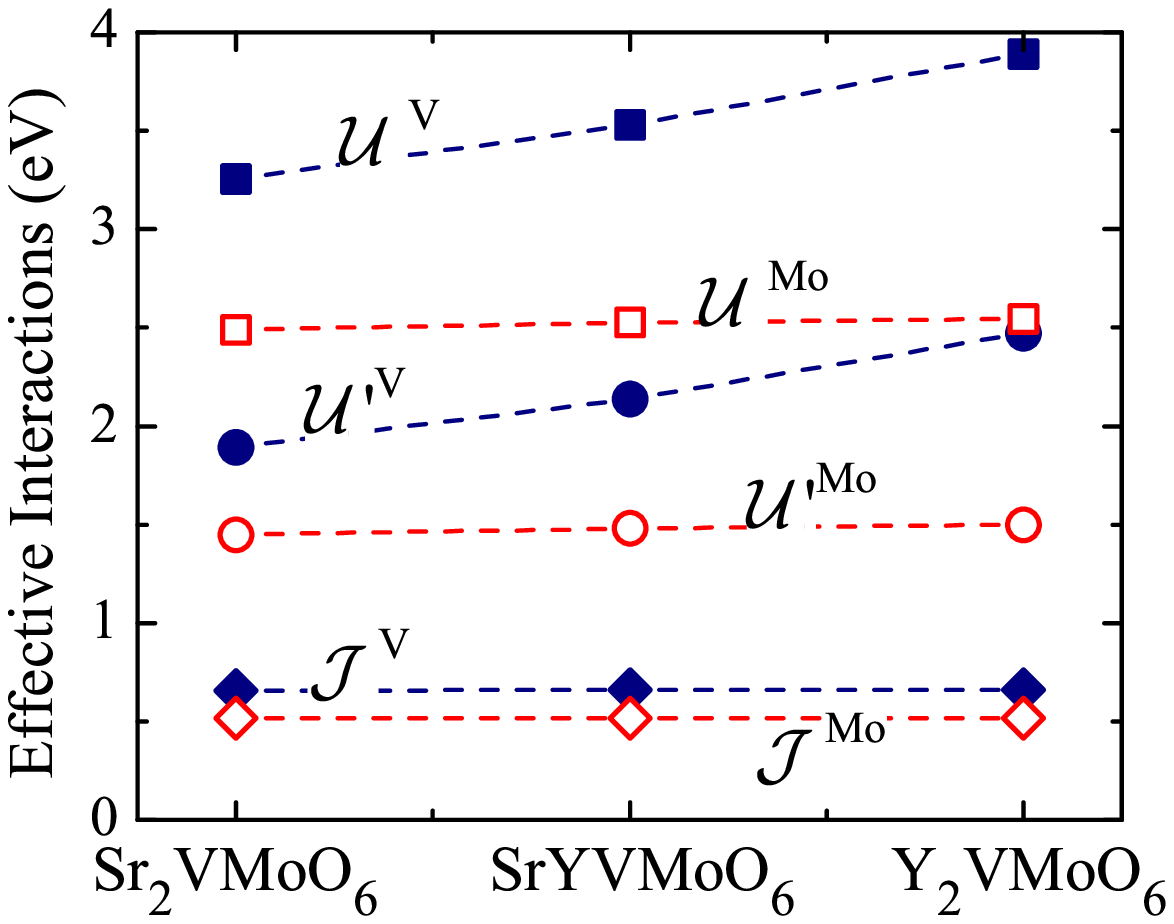}}
\resizebox{6cm}{!}{\includegraphics{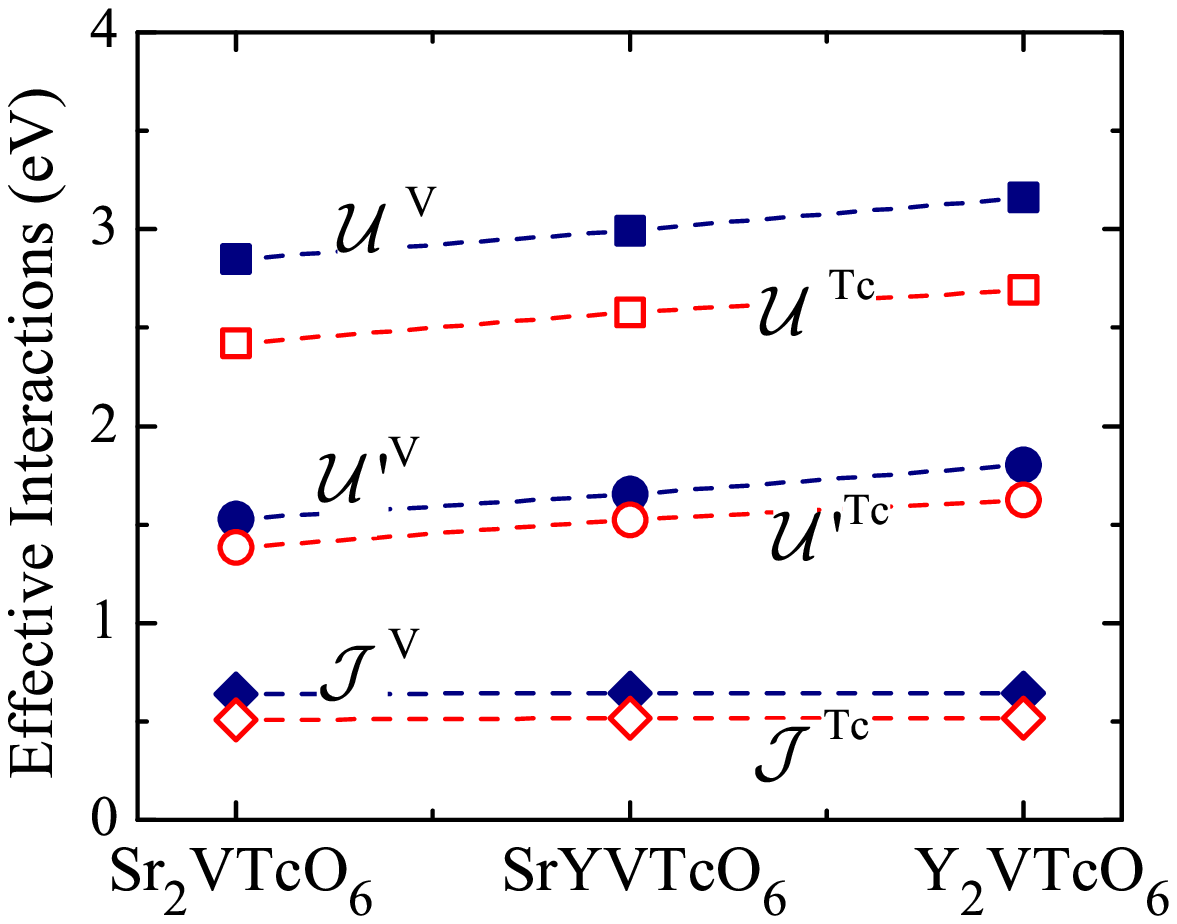}}
\end{center}
\caption{\label{fig.UJ} Intraorbital Coulomb interaction
$\cal{U}$, interorbital Coulomb interaction $\cal{U}'$, and the exchange
interaction $\cal{J}$ for the Sr$_{2-x}$Y$_x$VMoO$_6$ (left) and
Sr$_{2-x}$Y$_x$VTcO$_6$ (right) compounds.}
\end{figure}
Generally, as the number of $t_{2g}$-electrons increases, the effective Coulomb interactions
also increase, reflecting the tendencies of the RPA screening~\cite{review2008}.
On the other hand, the effective exchange interaction $\mathcal{J}$ is less
sensitive to the band filling. The Coulomb repulsion $\mathcal{U}^{\rm V}$
is larger for the VMo-compounds, due to the specific electronic
structure of Sr$_{2-x}$Y$_x$VMoO$_6$ (Fig.~\ref{fig.LDA}): since the V-band is separated
and located in the low-energy part of the spectrum, it will be populated
in the process of the band-filling. It will effectively reduce the number of channel available
for the RPA screening of $\mathcal{U}^{\rm V}$, for example, due to the transitions from the occupied
oxygen band to the unoccupied part of the V-band~\cite{review2008}. Thus, $\mathcal{U}^{\rm V}$
will be less screened in the VMo-systems in comparison with the VTc-ones, where the V- and Tc-states
form one common band (Fig.~\ref{fig.LDA}).
\end{enumerate}

\subsection{\label{sec:OEP} Optimized effective potential method}
  In order to solve the model and find the magnetic ground state of
the double perovskites,
we employ the optimized effective potential method.
The OEP model works in the spirit of the Kohn-Sham
density functional theory~\cite{KohnSham} on the double-perovskite lattice,
where the kinetic energy
$$
E_{\rm kin} = \sum_\sigma \sum_n^{\rm occ}  \frac{1}{\Omega_{\rm BZ}} \int d {\bf k}
\langle c_{\sigma n {\bf k}} | \hat{t}_{\bf k} | c_{\sigma n {\bf k}} \rangle,
$$
the number of electrons
$$
n^{\tau} = \sum_\sigma \sum_m \sum_n^{\rm occ} \frac{1}{\Omega_{\rm BZ}} \int d {\bf k}
c^{m \tau \dagger}_{\sigma n {\bf k}} c^{m \tau}_{\sigma n {\bf k}} ,
$$
and the spin magnetic moment
$$
m^{\tau} = \sum_m \sum_n^{\rm occ} \frac{1}{\Omega_{\rm BZ}} \int d {\bf k}
\left( c^{m \tau \dagger}_{\uparrow n {\bf k}} c^{m \tau}_{\uparrow n {\bf k}}
- c^{m \tau \dagger}_{\downarrow n {\bf k}} c^{m \tau}_{\downarrow n {\bf k}}
\right)
$$
at the site $\tau$ are constructed from the one-electron orbitals
$c_{\sigma n {\bf k}}$,
which are obtained from the solution of the
auxiliary one-electron problem
\begin{equation}
\left( \hat{t}_{\bf k} + \hat{v}_{\rm eff}^\sigma \right) | c_{\sigma n {\bf k}} \rangle =
\varepsilon_{\sigma n {\bf k}} | c_{\sigma n {\bf k}} \rangle
\label{eqn:KS}
\end{equation}
with the potential $\hat{v}_{\rm eff}^\sigma$.
In the above notations,
$\hat{t}_{\bf k}$ is the Fourier image of
$\hat{t}_{ij} = \| t_{ij}^{m m'} \|$,
each eigenvector $| c_{\sigma n {\bf k}} \rangle$ is specified
by its components $c^{m \tau}_{\sigma n {\bf k}}$
in the basis of $t_{2g}$-orbitals $m$$=$ $xy$, $yz$, or $zx$
of the atomic type
$\tau =$ V or Mo(Tc),
$n$ is the band index, and ${\bf k}$ is
the position in the first Brillouin zone with the volume $\Omega_{\rm BZ}$.

  The parameters of the potential $\hat{v}_{\rm eff}^\sigma$,
which is supposed to be ``local'' (or diagonal with respect to the site indices), are obtained
from the minimization of the total energy
\begin{equation}
E = E_{\rm kin} + E_{\rm C} + E_{\rm X} + E_{\rm corr},
\label{eqn:Etot}
\end{equation}
where
the Coulomb ($E_{\rm C}$) and exchange ($E_{\rm X}$) energies
are given by
$$
E_{\rm C} = \frac{1}{2} \sum_\tau \sum_{\sigma \sigma'} \sum_{m m' m'' m'''}
U^\tau_{m m' m'' m'''} n^{\sigma \tau}_{m m'} n^{\sigma' \tau}_{m'' m'''}
$$
and
$$
E_{\rm X} = -\frac{1}{2} \sum_\tau \sum_\sigma \sum_{m m' m'' m'''}
U^\tau_{m m''' m'' m'} n^{\sigma \tau}_{m m'} n^{\sigma \tau}_{m'' m'''},
$$
respectively, in terms of the density matrix:
$$
n^{\sigma \tau}_{m m'} = \sum_n^{\rm occ} \frac{1}{\Omega_{\rm BZ}} \int d {\bf k}
c^{m \tau \dagger}_{\sigma n {\bf k}} c^{m' \tau}_{\sigma n {\bf k}} .
$$

  The expression for the correlation energy,
which is treated in the random phase approximation, is given by~\cite{Pines,BarthHedin,FerdiPRL2002}:
\begin{equation}
\fl
E_{\rm corr} = -\frac{1}{4 \Omega_{\rm BZ}} \int d {\bf q}
{\rm Tr} \left\{
\ln \left[1-\hat{P}(i\omega,{\bf q})\hat{U}\right]
\left[1-\hat{U}\hat{P}(i\omega,{\bf q})\right]
+ 2 \hat{P}(i\omega,{\bf q})\hat{U}
\right\},
\label{eqn:Ecorr}
\end{equation}
where $\hat{U}$ are the diagonal matrices $\| U_{m m' m'' m'''}^\tau \| $
with respect to the site indices $\tau$,
$\hat{P} = \| P^{\tau \tau'}_{m m' m'' m'''} \| $
is the polarization in the imaginary frequency:
\begin{equation}
\fl
P^{\tau \tau'}_{m m' m'' m'''}(i\omega,{\bf q}) = \sum_\sigma \sum_n^{\rm occ} \sum_{n'}^{\rm unocc}
\frac{1}{\Omega_{\rm BZ}} \int d {\bf k}
\frac{2 ( \varepsilon_{\sigma n {\bf k}} - \varepsilon_{\sigma n' {\bf k}+{\bf q}} )}
{\omega^2 + ( \varepsilon_{\sigma n {\bf k}} - \varepsilon_{\sigma n' {\bf k}+{\bf q}} )^2}
c_{\sigma n' {\bf k}+{\bf q}}^{m \tau \dagger} c_{\sigma n {\bf k}}^{m' \tau}
c_{\sigma n {\bf k}}^{m'' \tau' \dagger} c_{\sigma n' {\bf k}+{\bf q}}^{m''' \tau'} ,
\label{eqn:Polarization}
\end{equation}
and matrix multiplication in (\ref{eqn:Ecorr}) implies the convolution over
two orbital indices: $(\hat{U}\hat{P})^{\tau \tau'}_{m m' m'' m'''} =
\sum_{l l'} U^\tau_{m m' l l'} P^{\tau \tau'}_{l l' m'' m'''}$.

  For the cubic systems (and neglecting the orbital polarization effects), the effective
potential is specified by three parameters:
$\hat{v}_{\rm eff}^{\uparrow {\rm V} } = - \Delta_{\rm ex}^{\rm V}/2$,
$\hat{v}_{\rm eff}^{\downarrow {\rm V} } =   \Delta_{\rm ex}^{\rm V}/2$,
$\hat{v}_{\rm eff}^{\uparrow {\rm Mo(Tc)} } = \Delta_{\rm ct} - \Delta_{\rm ex}^{\rm Mo(Tc)}/2$, and
$\hat{v}_{\rm eff}^{\downarrow {\rm Mo(Tc)} } = \Delta_{\rm ct} + \Delta_{\rm ex}^{\rm Mo(Tc)}/2$.
In the other words, $\Delta_{\rm ex}^{\rm V}$ and
$\Delta_{\rm ex}^{\rm Mo(Tc)}$ is the
exchange-correlation splitting between minority- and majority-spin states
at the sites V and Mo(Tc), respectively,
and $\Delta_{\rm ct}$ is the ``charge transfer energy'' between sites V and Mo(Tc).
All these parameters are obtained by minimizing
the total energy (\ref{eqn:Etot}).
In practice, first we calculate the total energy on some coarse mesh
of these parameters, identify the minimum, then continue the calculations around the minimum
by using finer mesh, and so on until reaching necessary accuracy for the total energy minimum
(less than 1 meV).

\section{\label{sec:results} Results and Discussions}
\subsection{\label{sec:DOS} Electronic structure and arbitrariness of the OEP procedure
for half-metallic and insulating states}

\begin{figure}[h!]
\begin{center}
\resizebox{5cm}{!}{\includegraphics{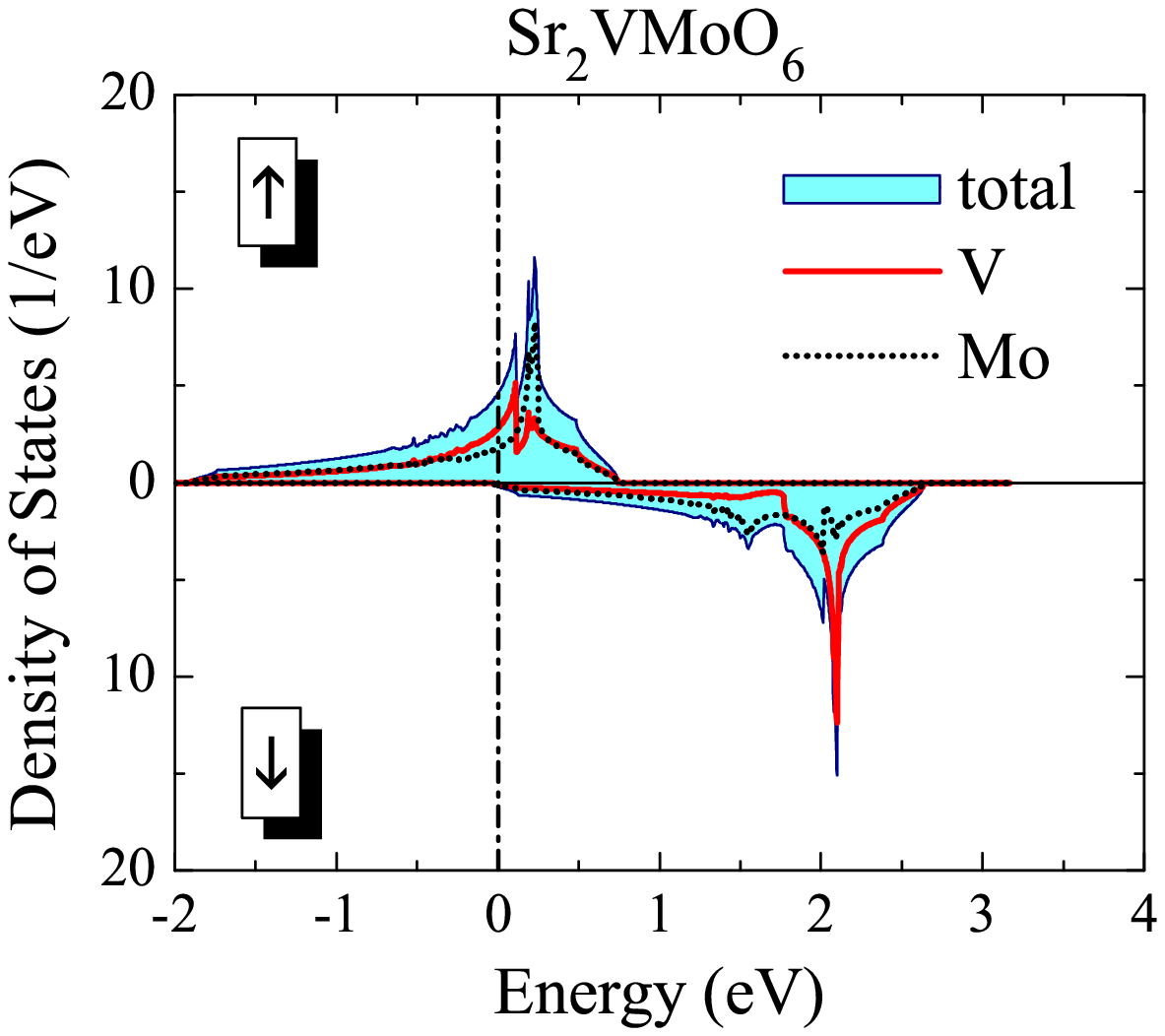}}
\resizebox{5cm}{!}{\includegraphics{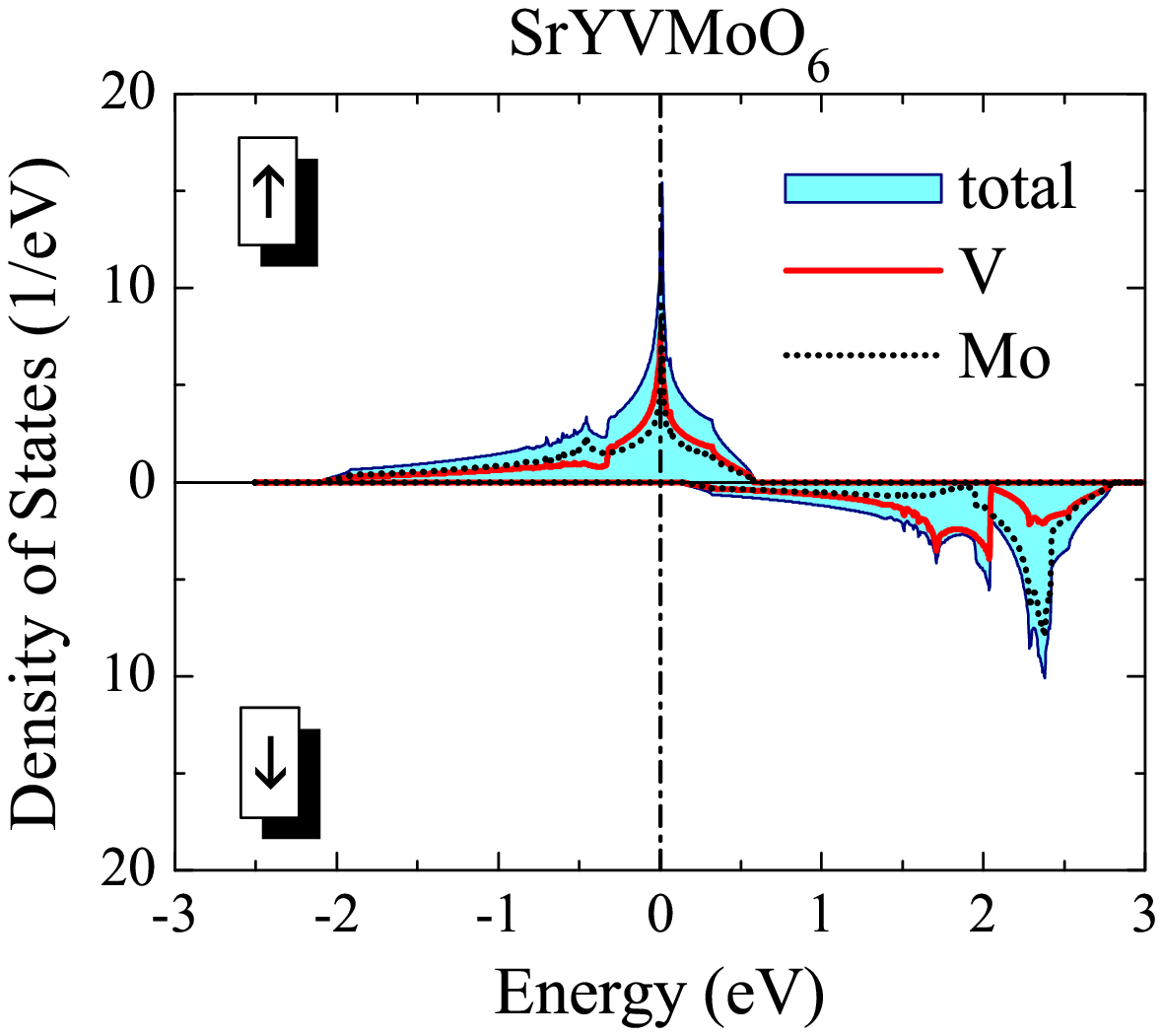}}
\resizebox{5cm}{!}{\includegraphics{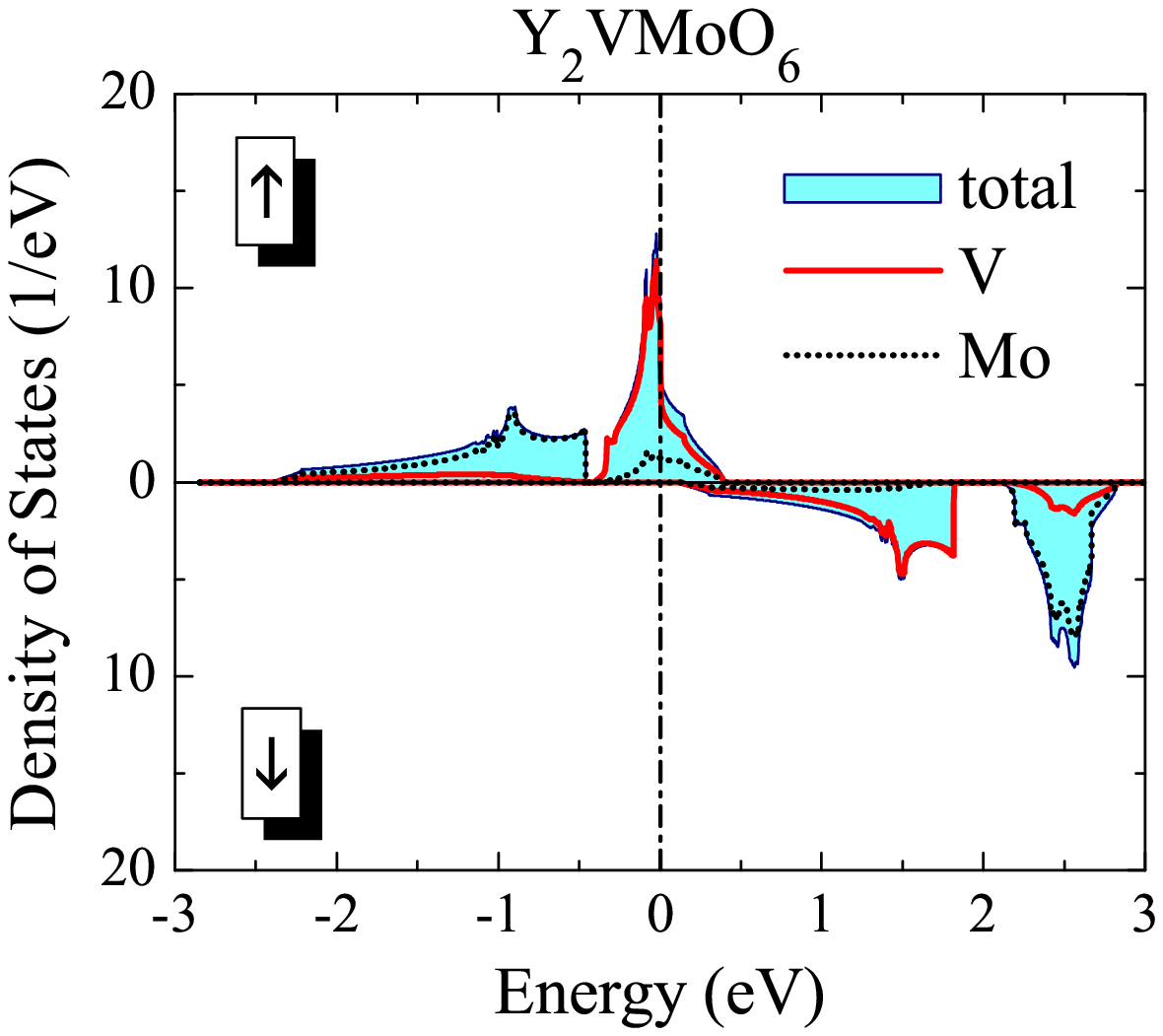}}
\end{center}
\begin{center}
\resizebox{5cm}{!}{\includegraphics{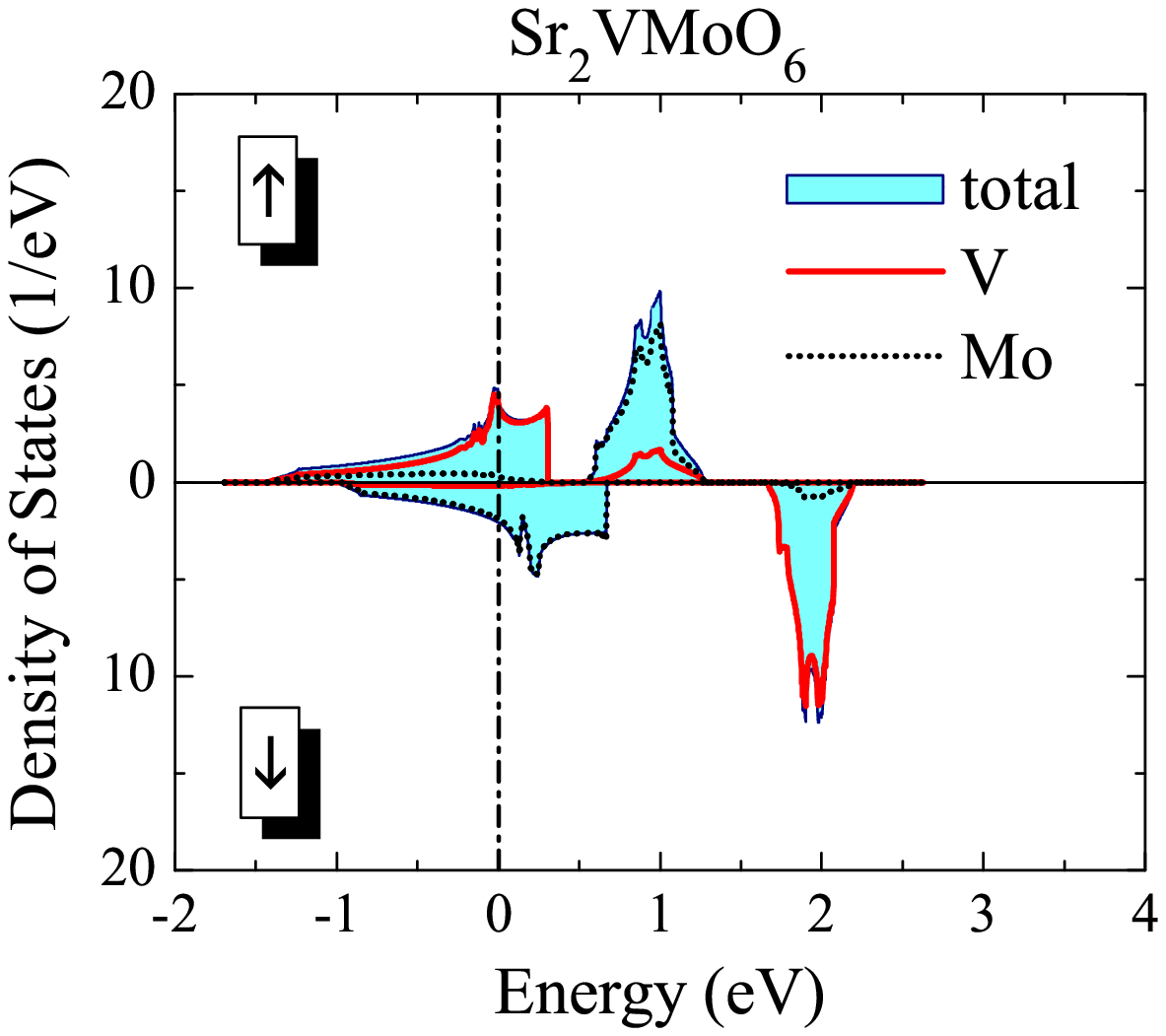}}
\resizebox{5cm}{!}{\includegraphics{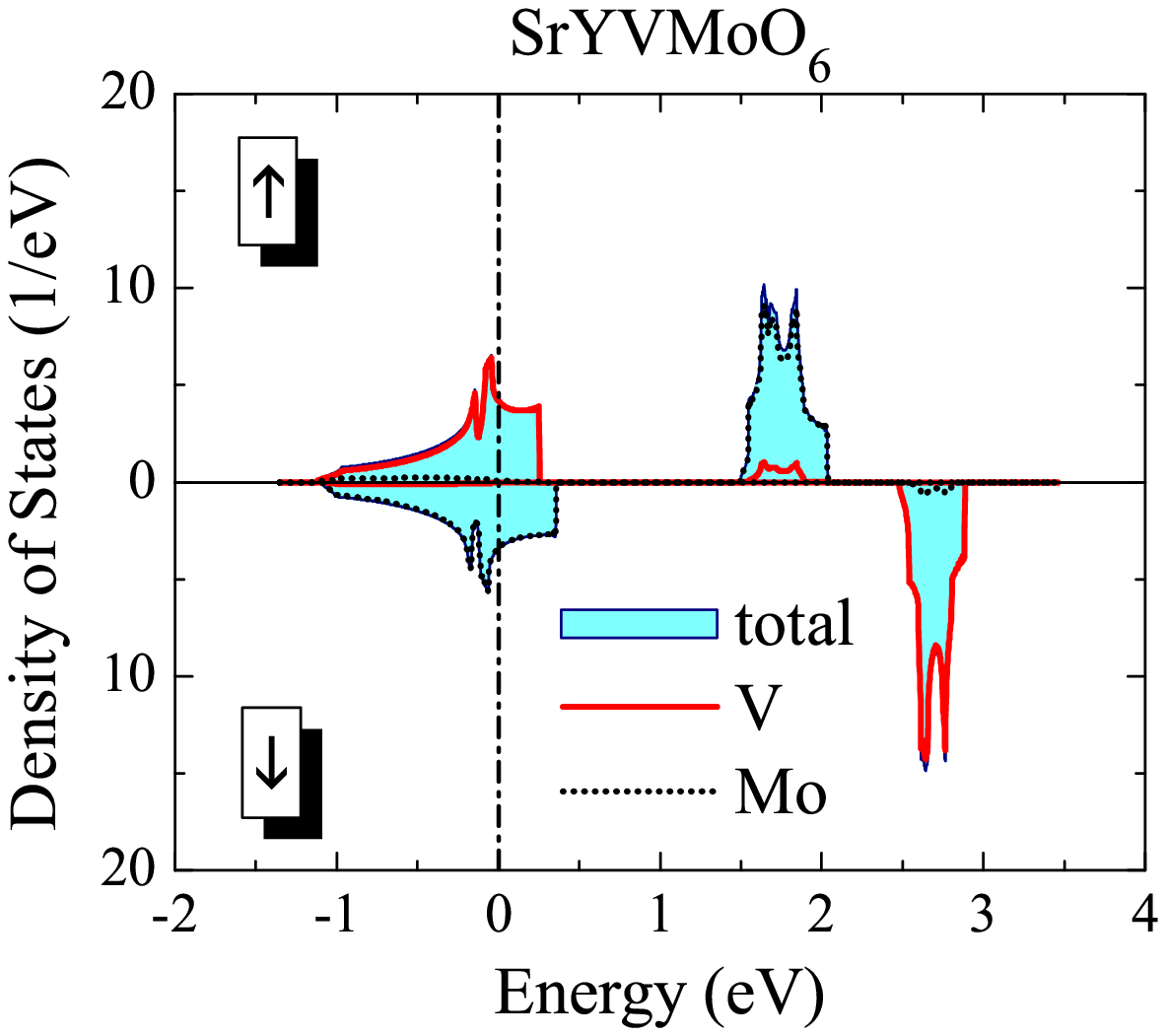}}
\resizebox{5cm}{!}{\includegraphics{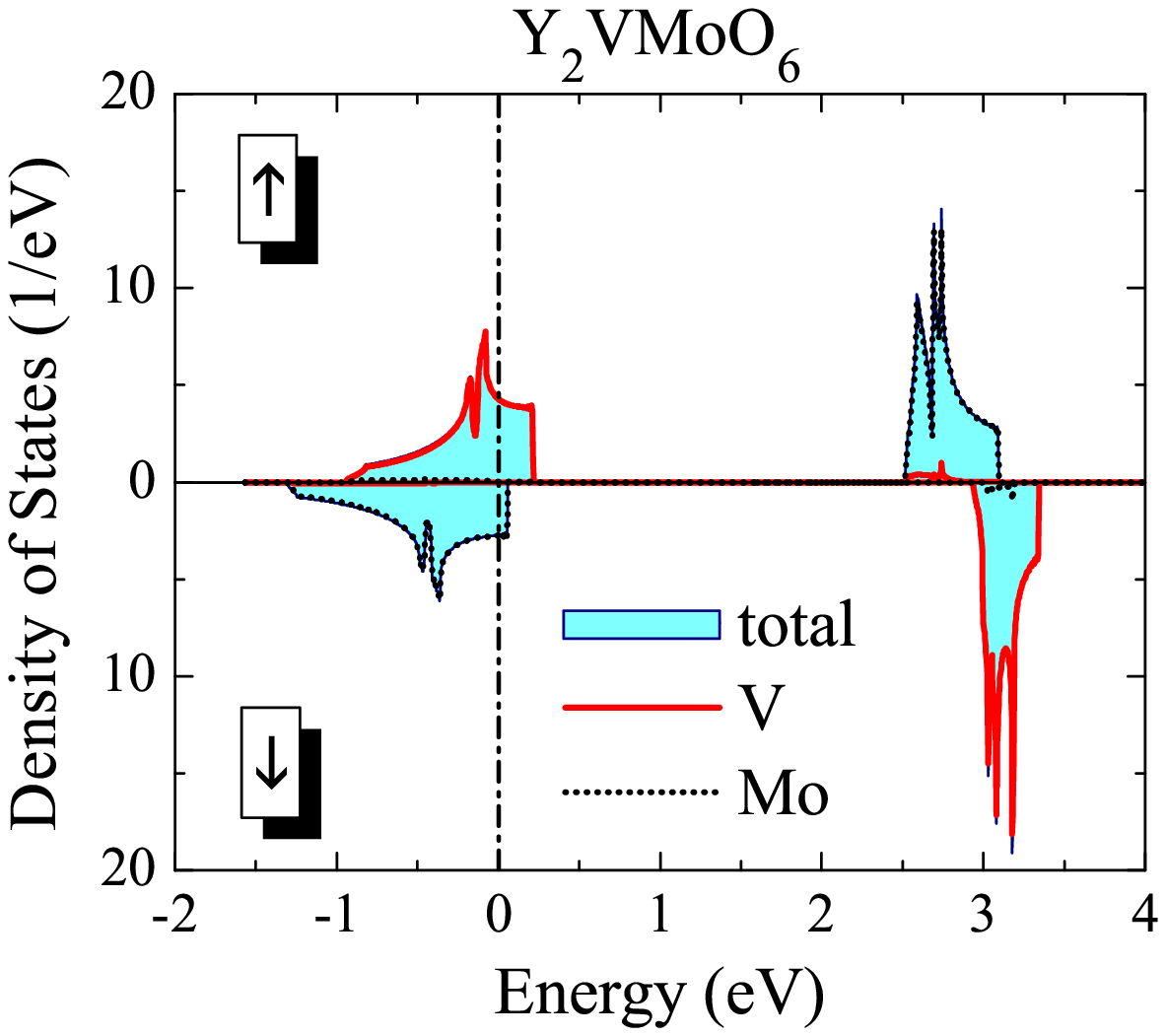}}
\end{center}
\caption{\label{fig.DOSVMo} Total and partial densities of states as
obtained
in the optimized effective potential
method
for the ferromagnetic (upper panel) and ferrimagnetic (lower panel)
configurations of Sr$_{2-x}$Y$_x$VMoO$_6$. The Fermi level is at zero energy.}
\end{figure}

  In the following, we consider two magnetic solutions:
ferromagnetic
($\Delta_{\rm ex}^{\rm V} > 0$ and $\Delta_{\rm ex}^{\rm Mo(Tc)} > 0$) and
ferrimagnetic
($\Delta_{\rm ex}^{\rm V} > 0$, while $\Delta_{\rm ex}^{\rm Mo(Tc)} < 0$),
and find parameters of the effective potential by minimizing the
total energy for each case. Total and partial densities of states
for Sr$_{2-x}$Y$_x$VMoO$_6$, obtained
from the solution of the Kohn-Sham equations (\ref{eqn:KS}) with the optimized potential,
are shown in Fig.~\ref{fig.DOSVMo}.
All FM solutions are half-metallic:
the Fermi levels crosses the majority-spin band, while the minority-spin band is
completely empty. In the case of
Sr$_2$VMoO$_6$ and SrYVMoO$_6$, there is a strong hybridization
between V- and Mo-states, which form the common majority-spin band. However, in Y$_2$VMoO$_6$
the majority-spin band
is split into the Mo- and V-subbands,
which are
separated by an energy gap.
This behavior is related to the large Coulomb repulsion $\mathcal{U}^{\rm V}$
in Y$_2$VMoO$_6$ (Fig.~\ref{fig.UJ}),
which additionally shifts
the V-states to the higher-energy part of spectrum. Any deviation from the
half-metallic behavior and partial population of the minority-spin band 
increases the total energy of Sr$_{2-x}$Y$_x$VMoO$_6$ and thus makes this state unstable.

  The choice of the parameters of the effective potential in the half-metallic regime
is not unique. For example, any redistribution of states in the unoccupied minority-spin
band does not affect the ground-state properties. Moreover,
any change of $\Delta_{\rm ex}^{\rm V}$, $\Delta_{\rm ex}^{\rm Mo}$,
and $\Delta_{\rm ct}$, which does not deform the partially populated
majority-spin band (apart from the constant shift), does not affect the ground-state
properties either. Quantitatively, this condition can be formulated
as follows: the ground-state properties in the half-metallic regime are controlled
by the single parameter
$a = \Delta_{\rm ct} + ( \Delta_{\rm ex}^{\rm V} - \Delta_{\rm ex}^{\rm Mo} )/2$.
Any change of $\Delta_{\rm ex}^{\rm V}$, $\Delta_{\rm ex}^{\rm Mo}$,
and $\Delta_{\rm ct}$, which keeps $a$, does not affect the ground-state properties
in the half-metallic regime.

  All ferrimagnetic configurations of Sr$_{2-x}$Y$_x$VMoO$_6$ are metallic.
In this case all three parameters
of the effective potential are uniquely defined. The degree of the spin polarization
($P$) in the ferrimagnetic state is sensitive to the concentration $x$ and the definition of $P$~\cite{Mazin}.
For example, using the simplest ``density of states'' definition
(see Fig.~\ref{fig.DOSVMo}), one obtains $P$$=$ 0.3, 0.1, and 0.2 for
$x$$=$ 0, 1 and 2, respectively. On the other hand, the ``Bloch-Bolzmann transport theory'' definition
yields $P$$=$ 0.1, 0.1, and 0.5 for $x$$=$ 0, 1 and 2, respectively.
The experimental value of $P$ for SrLaVMoO$_6$, measured by using the point-contact Andreev reflection technique,
is about 0.5~\cite{Gotoh}.
Nevertheless, as we will see below, many of these ferrimagnetic states appear to be unstable and
calculated spin polarization cannot be directly compared
with the experimental data.

\begin{figure}[h!]
\begin{center}
\resizebox{5cm}{!}{\includegraphics{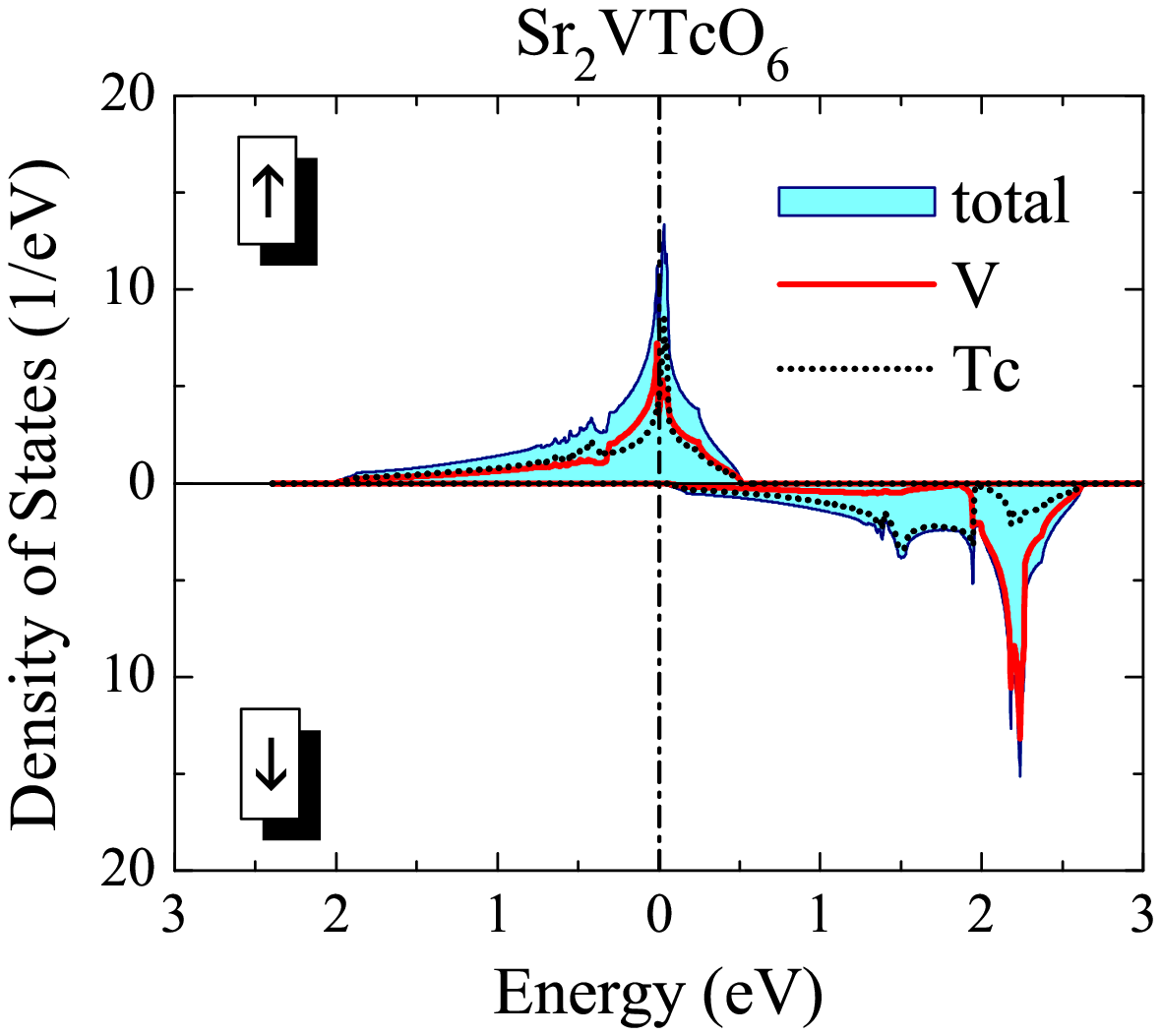}}
\resizebox{5cm}{!}{\includegraphics{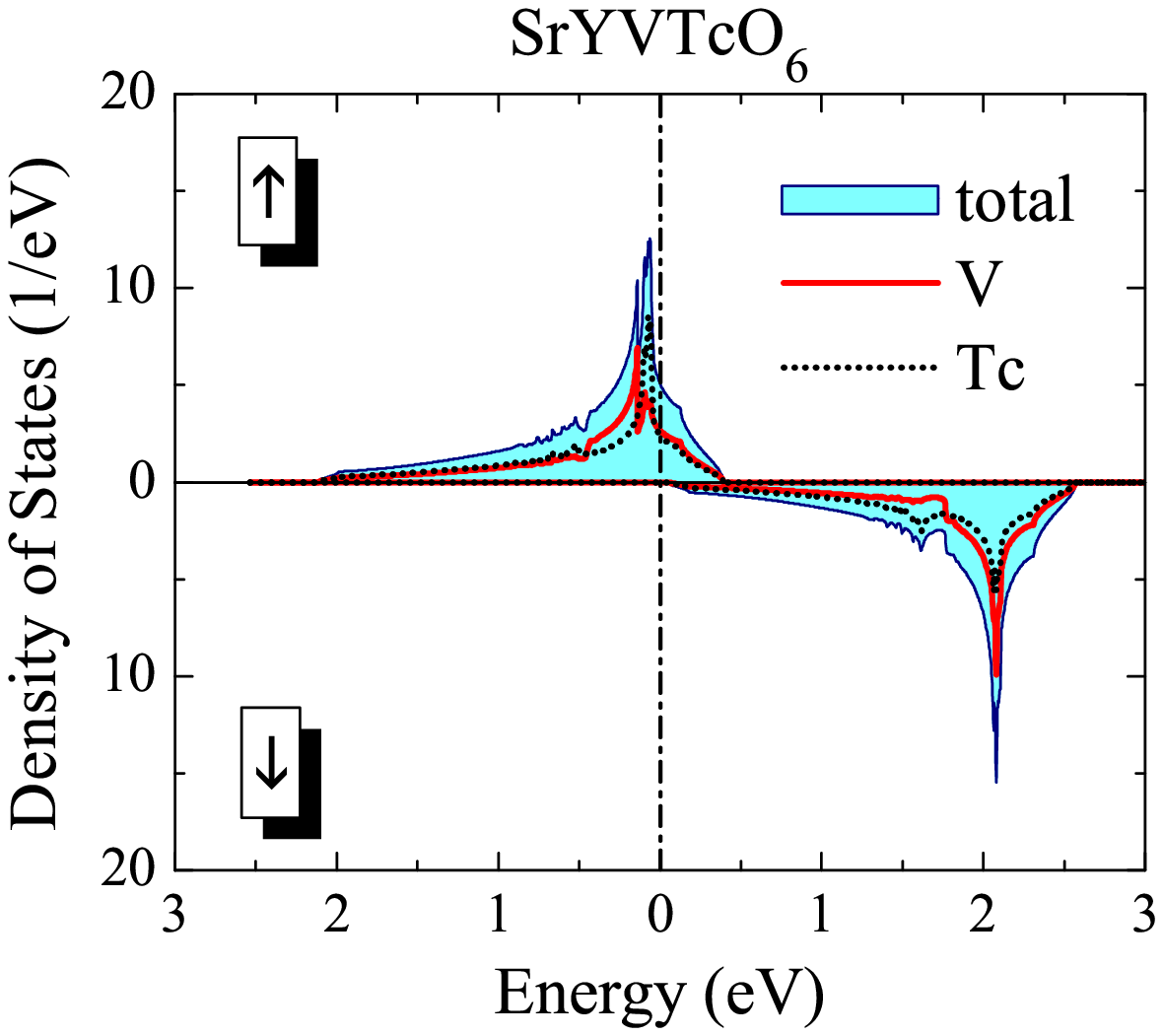}}
\end{center}
\begin{center}
\resizebox{5cm}{!}{\includegraphics{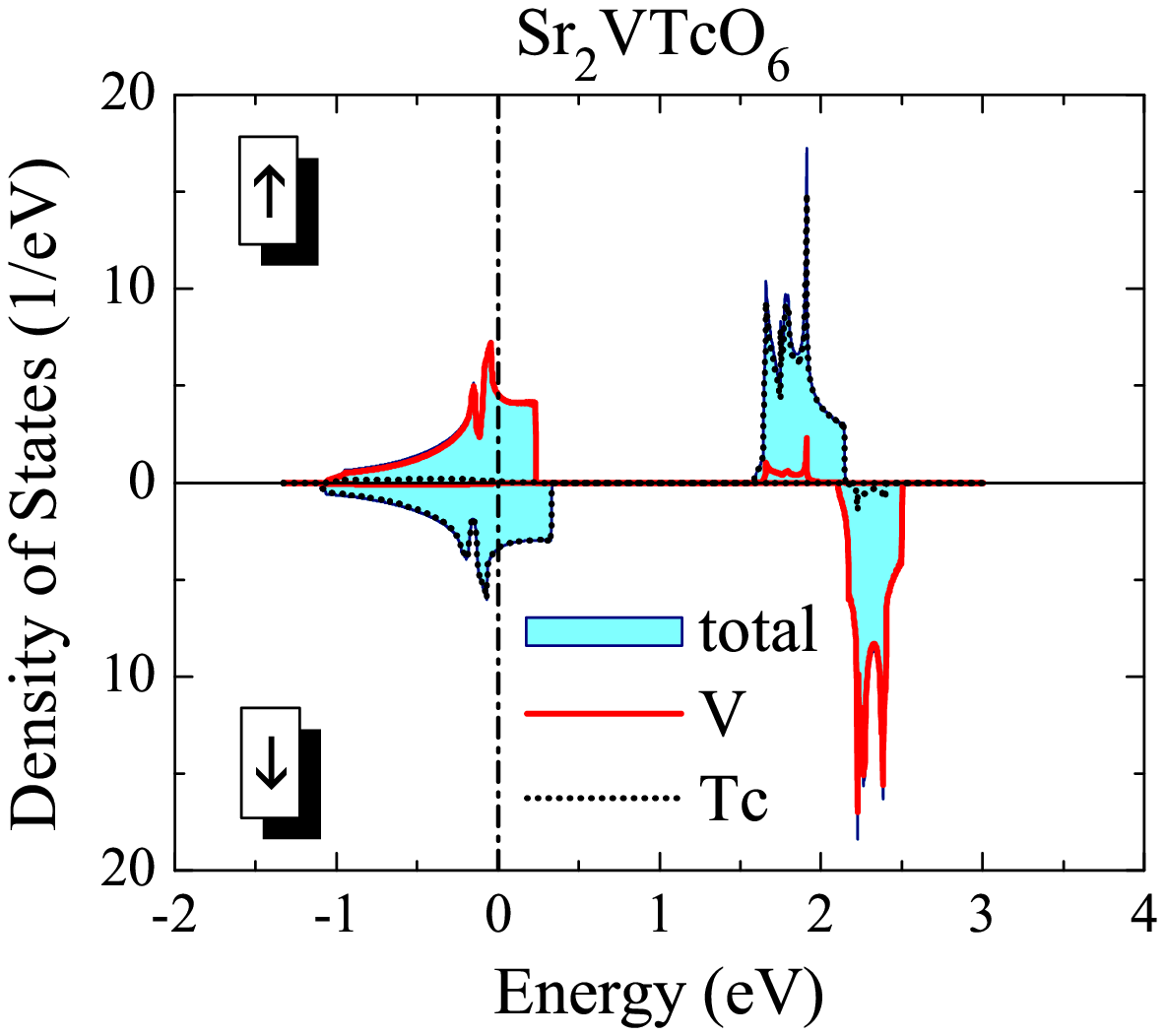}}
\resizebox{5cm}{!}{\includegraphics{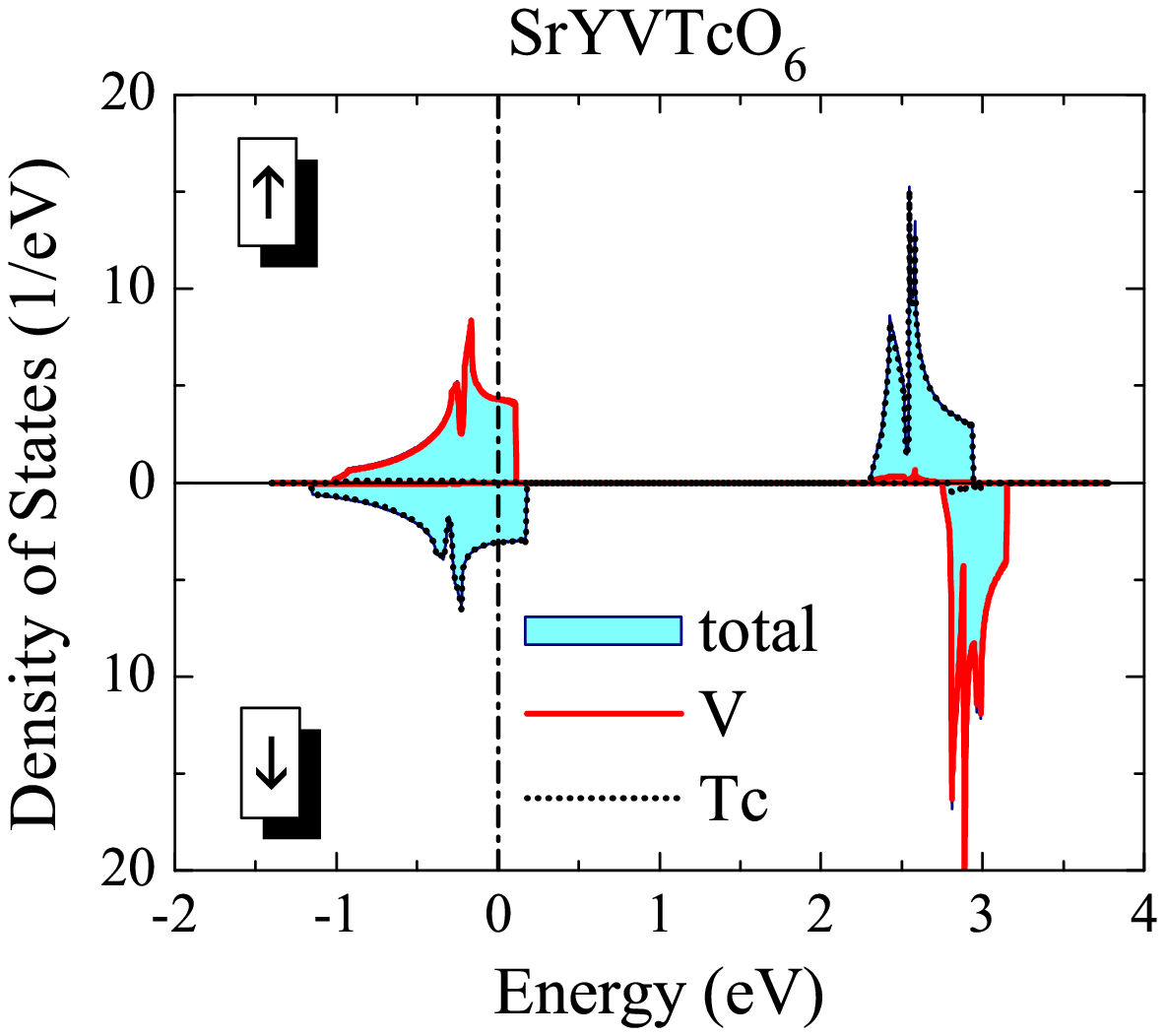}}
\end{center}
\caption{\label{fig.DOSVTc} Total and partial densities of states as
obtained
in the optimized effective potential
method
for the ferromagnetic (upper panel) and ferrimagnetic (lower panel)
configurations of Sr$_2$VTcO$_6$ and SrYVTcO$_6$.
The Fermi level is at zero energy.}
\end{figure}

  Similar behavior was obtained for Sr$_2$VTcO$_6$ and SrYVTcO$_6$, where
again half-metallic FM state competes with the normal metallic ferrimagnetic state (Fig.~\ref{fig.DOSVTc})

\begin{figure}[h!]
\begin{center}
\resizebox{5cm}{!}{\includegraphics{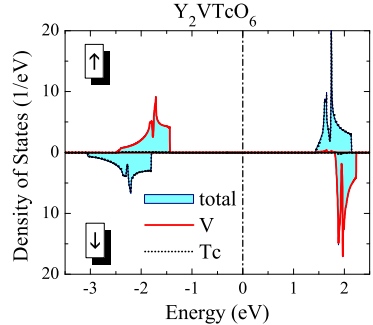}}
\end{center}
\caption{\label{fig.DOSY2VTc} Total and partial densities of states as
obtained
in the optimized effective potential
method
for the antiferromagnetic
configuration of Y$_2$VTcO$_6$. The Fermi level lies in the band gap.}
\end{figure}

  Y$_2$VTcO$_6$ requires a special attention: it has six electrons in the $t_{2g}$-band.
The system is insulating and all metallic configurations have higher energies.
Therefore, in the FM state, the parameters
$\Delta_{\rm ex}^{\rm V}$, $\Delta_{\rm ex}^{\rm Tc}$,
and $\Delta_{\rm ct}$ can be chosen in an arbitrary way.
Indeed, any combination of these parameters (provided that the
system remains insulating) yields the same number of electrons $n^{\tau} = 3$, 
spin magnetic
moments $m^{\tau} = 3 \mu_B$, and the total energy of the system. The correlation
energy is exactly equal to zero for this particular case.
The arbitrariness with the choice of the effective potential
remains also in the antiferromagnetic state of Y$_2$VTcO$_6$ (Fig.~\ref{fig.DOSY2VTc}),
although the number of arbitrary parameters decreases.
Indeed, the splitting between V- and Mo-states with the same spin affects
the ground-state properties. However, the bands with different spins can be
rigidly shifted relative to each other and, as long as the system remains in the
insulating state, this shift does not affect the ground-state properties.
Thus, the properties of the antiferromagnetic Y$_2$VTcO$_6$ are defined by only two
parameters: $\Delta_{\rm ct}$ and $( \Delta_{\rm ex}^{\rm V} - \Delta_{\rm ex}^{\rm Tc} )$,
while the third parameter $( \Delta_{\rm ex}^{\rm V} + \Delta_{\rm ex}^{\rm Tc} )$
can be taken arbitrarily. Finally, we would like to emphasize that
antiferromagnetic Y$_2$VTcO$_6$ becomes insulating due to the large exchange splitting
(both at the V- and Tc-sites) obtained in the OEP calculations. If the splitting were smaller,
the majority-spin V- and Tc-bands would eventually merge, yielding the half-metallic
antiferromagnetic state, which was obtained in the LSDA calculations for the isoelectronic
La$_2$VTcO$_6$~\cite{Wang}.
Thus, such half-metallic antiferromagnetic state is probably an
artifact of LSDA,
and more rigorous treatment of the correlation effects will open the band gap in
the both spin channels.

\subsection{\label{sec:StonerI} Screening and the effective Stoner model}

\begin{figure}[h!]
\begin{center}
\resizebox{5cm}{!}{\includegraphics{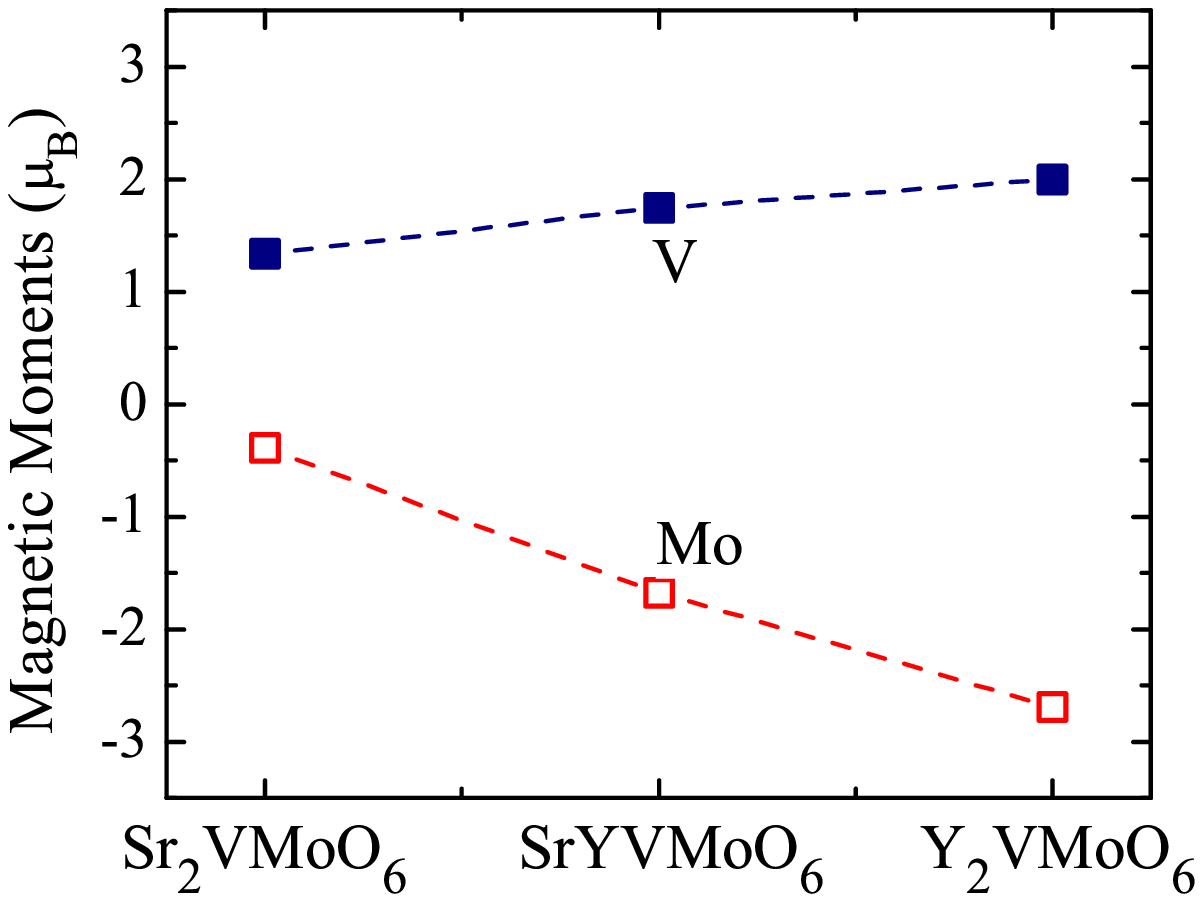}}
\resizebox{5cm}{!}{\includegraphics{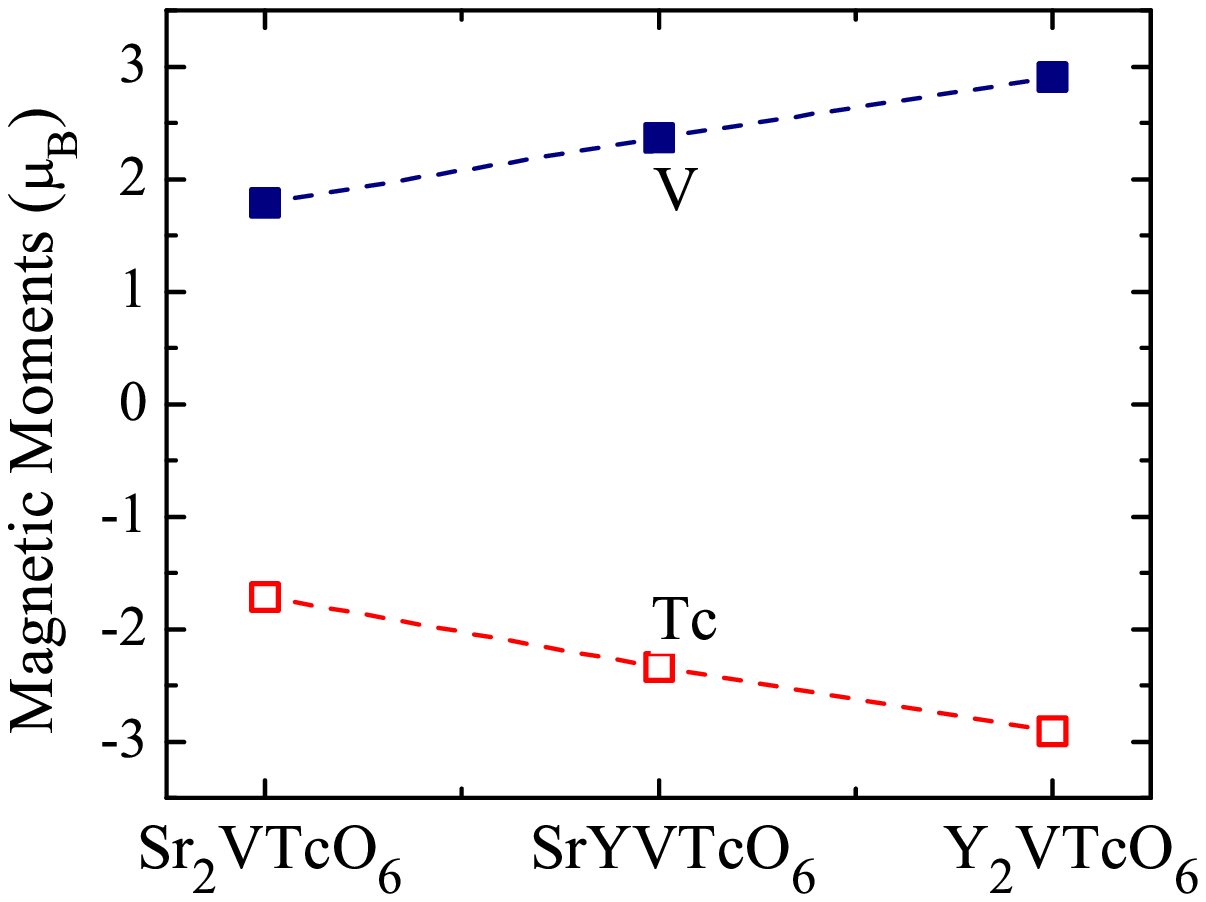}}
\end{center}
\caption{\label{fig.moments} Local spin magnetic moments
at the V- and Mo(Tc)-sites, as
obtained
in the optimized effective potential
method
for Sr$_{2-x}$Y$_x$VMoO$_6$ (left) and Sr$_{2-x}$Y$_x$VTcO$_6$ (right).}
\end{figure}

\begin{figure}[h!]
\begin{center}
\resizebox{10cm}{!}{\includegraphics{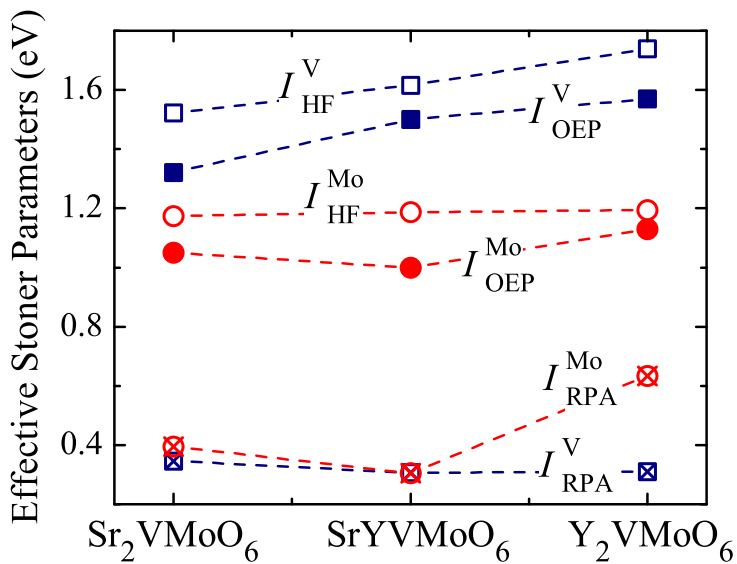}}
\end{center}
\caption{\label{fig.StonerVMo}
Effective Stoner parameters for
Sr$_{2-x}$Y$_x$VMoO$_6$ as derived in the optimized effective potential (OEP)
method for the ferrimagnetic state, in the Hartree-Fock (HF) approximation, and
using only static values of Coulomb and exchange interactions in the random phase approximation (RPA).}
\end{figure}

  Since the effective potential is local
(or diagonal with respect to the site indices)
and does not affect the orbital degrees of freedom, the OEP scheme has many similarities
with the Stoner model of magnetism, where the strength of the intraatomic
splitting $\Delta^\tau_{\rm ex}$
between the minority- and majority-spin states is controlled by the
effective parameter $I^\tau$:
$\Delta^\tau_{\rm ex} = I^\tau m^\tau$.
Thus, using the intraatomic exchange splitting,
obtained from the minimization of the total energy, and corresponding
values of the spin magnetic moments in the ground state (Fig.~\ref{fig.moments}),
one can find the effective Stoner parameter $I^\tau_{\rm OEP}$.
All such considerations are valid only
for the ferrimagnetic metallic state, where $\Delta^\tau_{\rm ex}$ are uniquely defined.
The results for
Sr$_{2-x}$Y$_x$VMoO$_6$ are explained in Fig.~\ref{fig.StonerVMo}.
The effective Stoner parameters derived from the OEP approach are of the order of 1.3-1.6 eV at the
V-sites and 1.0-1.1 eV at the Mo-sites. Similar parameters were obtained for
Sr$_{2-x}$Y$_x$VTcO$_6$: $I^{\rm V}_{\rm OEP}=1.27$ eV and
$I^{\rm Tc}_{\rm OEP}=1.06$ eV in the case of Sr$_2$VTcO$_6$, and
$I^{\rm V}_{\rm OEP}=1.28$ eV and $I^{\rm Tc}_{\rm OEP}=1.18$ eV in the case of SrYVTcO$_6$.
All of them are
systematically
larger that the typical values of $I^\tau$ obtained in the local-spin-density
approximation (typically, close 1 eV for the $3d$-elements and even smaller for the
$4d$- and $5d$-elements~\cite{Gunnarsson}).
This example clearly shows the importance of the rigorous treatment of
the exchange-correlation interactions (beyond LSDA)
in the physics of double perovskites.

  Another limiting case is the Hartree-Fock (HF) approximation,
which neglects the correlation interactions. Then,
corresponding Stoner parameter describing the splitting in the system of
three $t_{2g}$-orbitals, are given by
$I^\tau_{\rm HF} = (\mathcal{U}^\tau +2 \mathcal{J}^\tau )/3$.
They are also plotted in Fig.~\ref{fig.StonerVMo}.
The difference between $I^\tau_{\rm HF}$ and $I^\tau_{\rm OEP}$
is the measure of correlation interactions and their
impact on the magnetic properties of the double perovskites.
One can see that $I^\tau_{\rm OEP}$ is reduced by about 10-20\% in comparison
with $I^\tau_{\rm HF}$, which indicates the importance of correlation interactions.

  Finally, it is instructive to consider the
effect of the
static screening alone. The corresponding on-site interaction parameters
in RPA are given by
$$
\hat{U}_{\rm RPA} = \frac{1}{\Omega_{\rm BZ}} \int d {\bf q}
\left[1 - \hat{U} \hat{P}(0,{\bf q})\right]^{-1} \hat{U} ,
$$
which yields the Stoner parameter
$I^\tau_{\rm RPA} = (\mathcal{U}_{\rm RPA}^\tau +2 \mathcal{J}_{\rm RPA}^\tau )/3$.
Such a picture can be regarded as the static limit of the
$GW$ approximation~\cite{Hedin,FerdiGunnarsson}, where the
frequency-dependent screened Coulomb interaction
$\hat{U}_{\rm RPA}(i \omega)$ is replaced by its static counterpart
$\hat{U}_{\rm RPA} \equiv \hat{U}_{\rm RPA}(0)$~\cite{Ferdi04}.
Typically, it is sufficient to consider only site-diagonal
contributions to $\hat{U}_{\rm RPA}$, because if $\hat{U}$ is
site-diagonal, the off-diagonal contributions to $\hat{U}_{\rm RPA}$
are expected to be small~\cite{PRB06a}.
These results are also shown in Fig.~\ref{fig.StonerVMo}.
One can clearly see that the parameters $I^\tau_{\rm RPA}$ are overscreened
and strongly underestimated in comparison with $I^\tau_{\rm OEP}$.
This indicates the importance of
dynamical screening effects in the construction of the
optimized effective potential.

\subsection{\label{sec:Etot} Total energy and stability of the ferrimagnetic state}

\begin{figure}[h!]
\begin{center}
\resizebox{8cm}{!}{\includegraphics{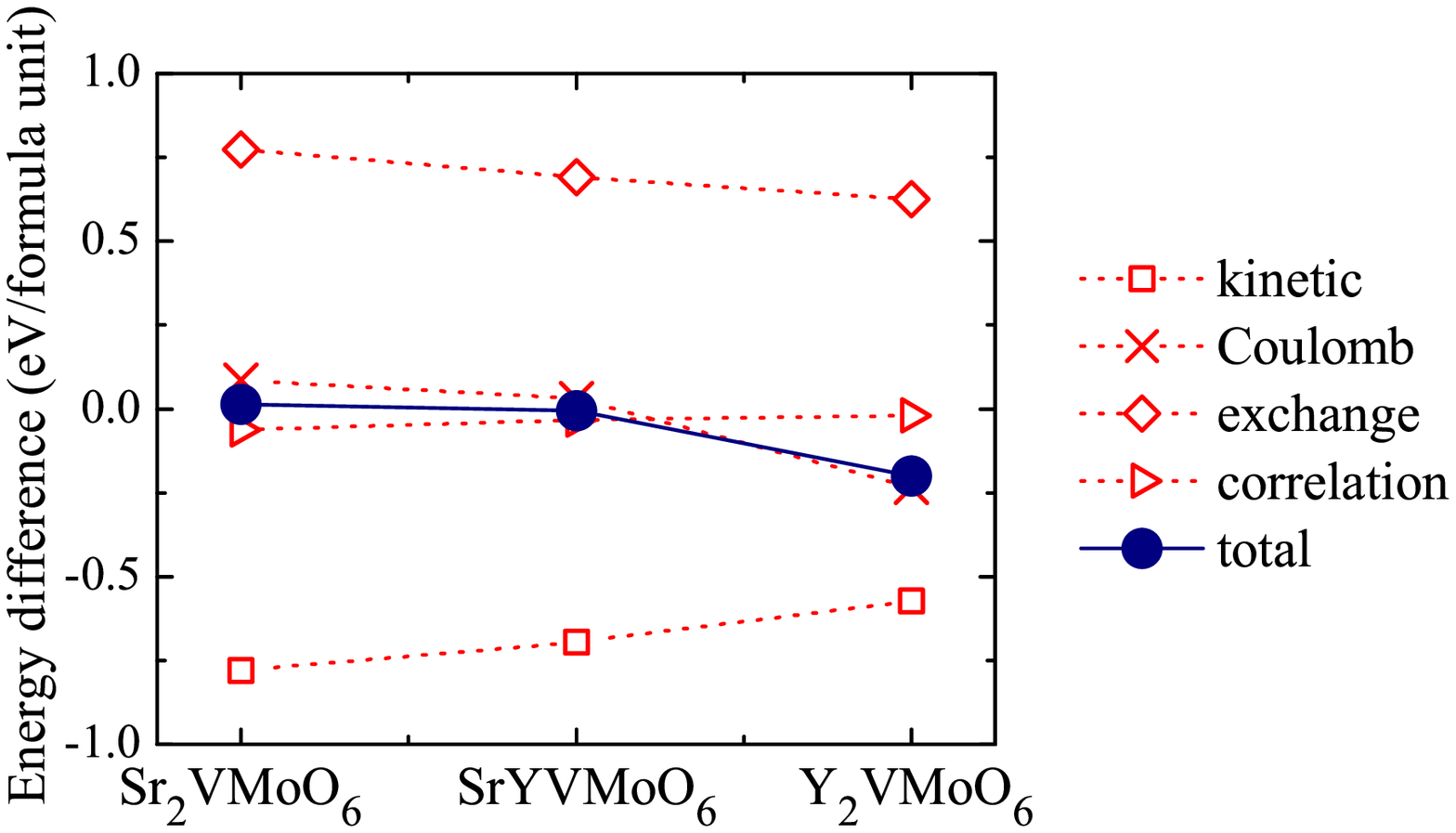}}
\end{center}
\begin{center}
\resizebox{8cm}{!}{\includegraphics{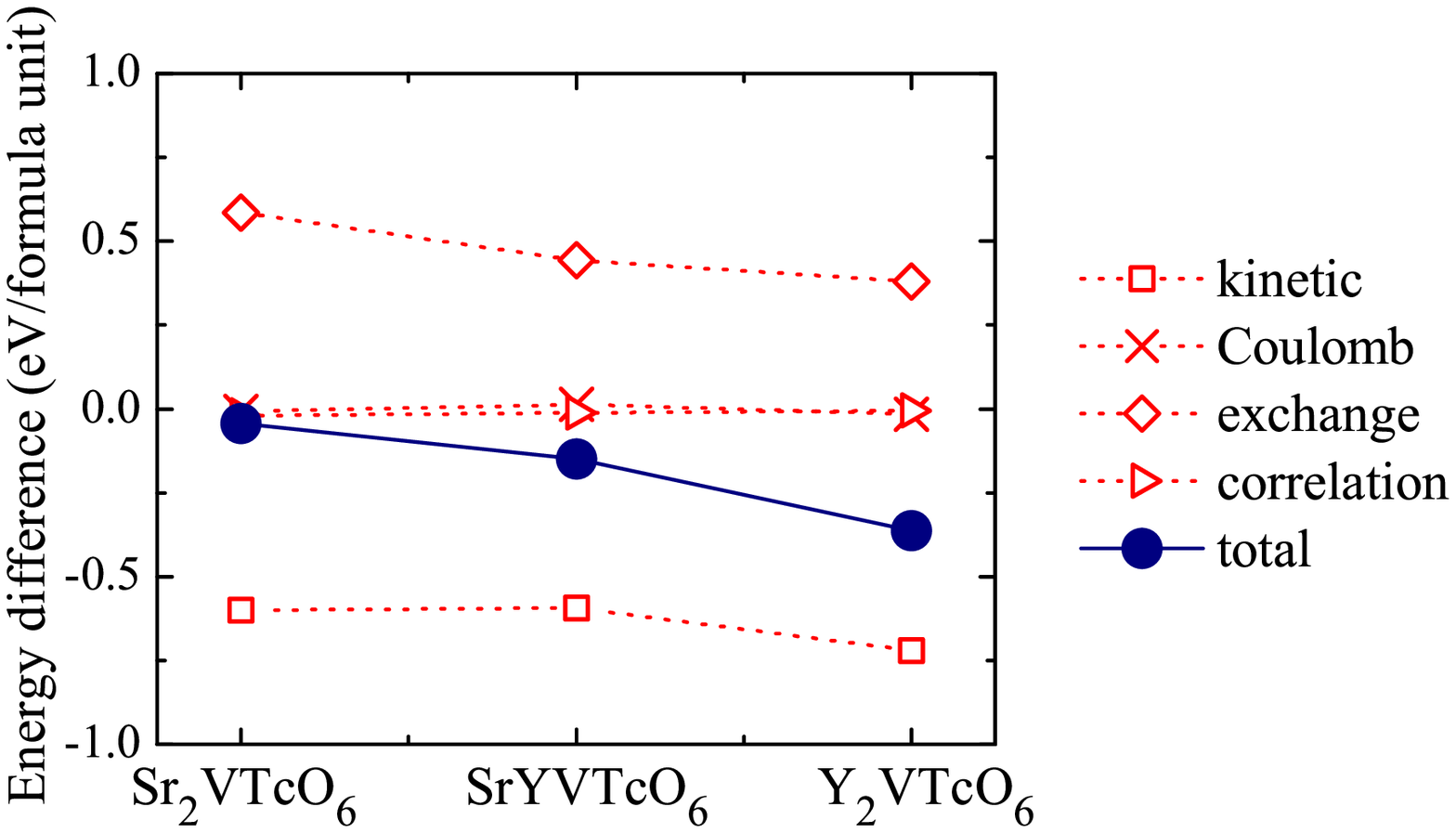}}
\end{center}
\caption{\label{fig.Etot} Total energy difference between ferrimagnetic
and ferromagnetic states and partial contributions to this energy difference,
calculated
for Sr$_{2-x}$Y$_x$VMoO$_6$ (top) and Sr$_{2-x}$Y$_x$VTcO$_6$ (bottom). }
\end{figure}

  The behavior of the total energy and of its partial contributions
is explained in Fig.~\ref{fig.Etot}.
For all considered compounds, except Sr$_2$VMoO$_6$, the ferrimagnetic configuration
has lower energy. The tendencies towards ferrimagnetism increase
with the band-filling.
The main factor stabilizing the ferrimagnetic state is the gain of the
kinetic energy. The main factor destabilizing it is the loss of the exchange energy.

  In order to investigate the local stability of the ferrimagnetic phase, we calculate
the spin stiffness $D$, which is related to the second derivative of the total energy
with respect to the spin-spiral vector ${\bf q}$ in the ferrimagnetic state:
$$
E({\bf q}) \approx E({\bf q}_0) + D ({\bf q} - {\bf q}_0)^2,
$$
where ${\bf q}_0 = (\pi/a,\pi/a,\pi/a)$ and $a$ is the cubic lattice parameter.
More specifically, the numerical calculations of $E({\bf q})$,
were performed by employing the generalized Bloch theorem~\cite{Sandratskii}
and the so-called local-force theorem for the infinitesimal spin rotations.
The latter states that the total energy change for the small rotations can be
replaced by the change of the one-electron energies, obtained from the
Kohn-Sham equations~(\ref{eqn:KS}) for the rigid spin rotations
of the effective potential $\hat{v}_{\rm eff}^\sigma$~\cite{Bruno}. The theorem
is rather general and
can be proven by considering the rotational invariance
of the exchange-correlation functions~\cite{PRB98}.
Thus, the range of the applicability of this theorem is not limited
by the local-spin-density approximation, for which it was originally proven,
but can be also extended to the OEP approach~\cite{PRB98}, provided that the effective potential
$\hat{v}_{\rm eff}^\sigma$ is uniquely defined.
Particularly,
since the choice of $\hat{v}_{\rm eff}^\sigma$ is not unique
in the case of insulating Y$_2$VTcO$_6$, this strategy cannot be directly applied to this system.

\begin{figure}[h!]
\begin{center}
\resizebox{10cm}{!}{\includegraphics{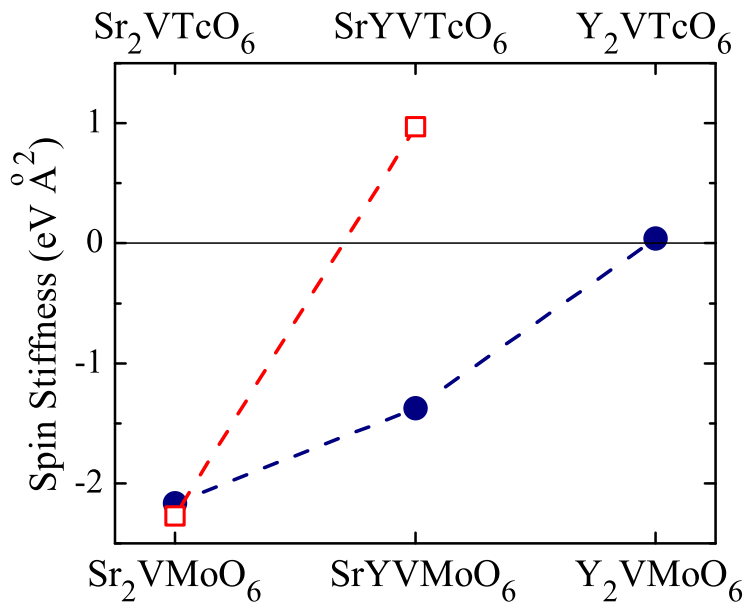}}
\end{center}
\caption{\label{fig.Stiffness} Spin stiffness in the ferrimagnetic state
calculated using optimized effective potentials
for Sr$_{2-x}$Y$_x$VMoO$_6$ (filled circles) Sr$_{2-x}$Y$_x$VTcO$_6$ (empty squares). }
\end{figure}

  The behavior of
$D$ in the OEP model is explained in Fig.~\ref{fig.Stiffness}.
Particularly, $D$ is negative in the case of Sr$_2$VMoO$_6$,
Sr$_2$VTcO$_6$, and SrYVMoO$_6$. Thus, the ferrimagnetic state in these compounds is
unstable with respect to the incommensurate spin-spiral state.
Since FM state has lower energy in the case of Sr$_2$VMoO$_6$ (Fig.~\ref{fig.Etot}),
the spin-spiral instability
of the ferrimagnetic state may
be finally resolved in the favor of the
collinear FM arrangement, to make it
the true magnetic ground state of Sr$_2$VMoO$_6$. However, in
Sr$_2$VTcO$_6$ and SrYVTcO$_6$ already ferrimagnetic state has lower energy. 
Moreover, this ferrimagnetic state itself appears to be unstable with respect to the
noncollinear
spin-spiral alignment. Thus, the ground state in this case will be
neither ferro- nor ferrimagnetic.

\section{\label{sec:summary} Summary and Conclusions}

  In summary, using first-principles electronic structure calculations,
we have constructed the effective low-energy model for the $t_{2g}$-states of
double perovskites Sr$_{2-x}$Y$_x$VMoO$_6$ and Sr$_{2-x}$Y$_x$VTcO$_6$.
Then, the ground-state properties of the obtained model
have been investigated on the basis of the
optimized effective potential method, by treating correlation interactions in the
random-phase approximation.
The reason for using such strategy was twofold.
On the one hand, such approach allows one to improve
the commonly used 
local-spin-density
approximation for the double perovskites and incorporate the physics of electron correlations
beyond the homogeneous electron gas limit. On the other hand, since OEP is formulated
as the ground-state method, where all the parameters are derived rigorously by minimizing the total energy of the
system,
it gets rid of many ambiguities
inherent to the LSDA$+$$U$ method.
Particularly, although Sr$_{2-x}$Y$_x$VMoO$_6$, Sr$_{2-x}$Y$_x$VTcO$_6$, and other isoelectronic to them compounds
are frequently proposed to exhibit half-metallic ferromagnetic or antiferromagnetic
properties, the situation is probably not such optimistic
because of the electron correlation effects, which affect the intraatomic
exchange splitting and makes it substantially larger than in LSDA.
Of course, this will revise many predictions based on the LSDA. For example, Y$_2$VTcO$_6$, which was
expected to be half-metallic antiferromagnet, appears to be insulator.
The behavior of other materials is even more complicated. For the most of them (except Sr$_2$VMoO$_6$),
the half-metallic ferromagnetic state is energetically less favorable than the (normal) metallic
ferrimagnetic one. Nevertheless, this ferrimagnetic state
appears to be also unstable with respect to the spin-spiral alignment. Thus, the true magnetic ground state of these
double perovskites will be the spin spiral or some more complicated non-collinear magnetic structure.

  Finally, we have argued that the effective potential (and therefore the
one-electron band structure) for the half-metallic
and insulating
ferro- or antiferromagnetic state is not uniquely defined. Generally, there is a bunch of parameters, which yield the
same total energy, occupation numbers and spin magnetic moments at the transition-metal sites.

\ack
I wish to thank the support
from the Federal Agency for Science and Innovations, grant No. 02.740.11.0217,
during my stay at the Ural State Technical University (Ekaterinburg, Russia).

\end{document}